\documentclass[twocolumn]{autart}

\usepackage{graphicx}
\usepackage[compress]{natbib}
\usepackage{enumerate}
\usepackage{mathtools,amsmath,amssymb}
\usepackage{etoolbox}
\AtEndEnvironment{pf}{\null\hfill$\square$}%
\usepackage{xcolor}
\definecolor{oiBlue}{HTML}{0072B2}
\definecolor{oiOrange}{HTML}{E69F00}
\usepackage{wrapfig}

\usepackage{aligned-overset}

\usepackage[hidelinks]{hyperref}
\edef\endfrontmatter{\unexpanded\expandafter{\endfrontmatter}\noexpand\endNoHyper}
\usepackage{booktabs}
\usepackage{amsfonts}
\usepackage{nicefrac}
\usepackage{multirow}
\usepackage{makecell}

\usepackage{algorithm}
\usepackage{algpseudocode}

\usepackage{siunitx}
\usepackage[labelformat=simple]{subfig}

\usepackage{pgfplots}
\pgfplotsset{compat=newest}
\usetikzlibrary{
    arrows.meta,
    decorations.markings,
}

\allowdisplaybreaks

\DeclareMathOperator*{\argmax}{arg\,max}
\DeclareMathOperator*{\argmin}{arg\,min}

\newcommand{\R}{\mathbb{R}}

\newcommand{\Prob}{\mathbb{P}}
\newcommand{\Expect}{\mathbb{E}}
\newcommand{\N}{\mathcal{N}}

\newcommand{\sph}{\mathbb{S}}

\newcommand{\subG}[1]{\mathrm{subG}({#1})}
\newcommand{\subExp}[2]{\mathrm{subExp}({#1}, {#2})}
\newcommand{\subGvec}[2]{\mathrm{subG}_{#2}({#1})}
\newcommand{\uniform}[1]{U(#1)}
\newcommand{\bigO}{\mathcal{O}}
\newcommand{\simiid}{\stackrel{\text{i.i.d.}}{\sim}}

\newcommand{\SOS}{\mathrm{SOS}}

\usepackage{acronym}
\acrodef{OLS}{ordinary least squares}
\acrodef{LTI}{linear time-invariant}
\acrodef{LMI}{linear matrix inequality}
\acrodefplural{LMI}{linear matrix inequalities}
\acrodef{SNR}{signal-to-noise ratio}
\acrodef{RoA}{region of attraction}
\acrodefplural{RoA}{region of attractions}
\acrodef{SOS}{sum-of-squares}
\acrodef{ISS}{input-to-state stability}
\acrodef{SDP}{semi-definite program}

\renewenvironment{thebibliography}[1]{%
    \begin{oldthebibliography}{#1}%
    \setlength{\parskip}{0ex}%
    \setlength{\itemsep}{0ex}%
}%
{%
    \end{oldthebibliography}%
}
\begin{document}
\setlength{\abovedisplayshortskip}{0.6ex plus1ex minus1ex}
\setlength{\abovedisplayskip}{0.6ex plus1ex minus1ex}
\setlength{\belowdisplayshortskip}{0.9ex plus1ex minus1ex}
\setlength{\belowdisplayskip}{0.9ex plus1ex minus1ex}
\begin{frontmatter}

\title{End-to-end guarantees for indirect data-driven control of bilinear systems with finite stochastic data \thanksref{footnoteinfo}}

\thanks[footnoteinfo]{
    This work is funded by Deutsche Forschungsgemeinschaft (DFG, German Research Foundation) under Germany's Excellence Strategy -- EXC 2075 -- 390740016 and within grant AL 316/15-1 -- 468094890. We acknowledge the support by the Stuttgart Center for Simulation Science (SimTech).
    N.\ Chatzikiriakos and R.\ Strässer thank the Graduate Academy of the SC SimTech for its support.
    \\
    \phantom{11}\emph{Email address:}
    \texttt{chatzikiriakos@ist.uni-stuttgart.de} (N. Chatzikiriakos),
    \texttt{straesser@ist.uni-stuttgart.de} (R. Strässer),
    \texttt{allgower@ist.uni-stuttgart.de} (F. Allgöwer),
    \texttt{iannelli@ist.uni-stuttgart.de} (A. Iannelli).
}

\author{Nicolas Chatzikiriakos}, 
\author{Robin Str\"asser}, 
\author{Frank Allg\"ower}, 
\author{Andrea Iannelli}

\address{
    University of Stuttgart, Institute for Systems Theory and Automatic Control, 70550 Stuttgart, Germany
}

\begin{abstract}
    In this paper we propose an end-to-end algorithm for indirect data-driven control for bilinear systems with stability guarantees.
    We consider the case where the collected i.\,i.\,d. data is affected by probabilistic noise with possibly unbounded support and leverage tools from statistical learning theory to derive finite sample identification error bounds. 
    To this end, we solve the bilinear identification problem by solving a set of linear and affine identification problems, by a particular choice of a control input during the data collection phase. 
    We provide a priori as well as data-dependent finite sample identification error bounds on the individual matrices as well as ellipsoidal bounds, both of which are structurally suitable for control. 
    Further, we integrate the structure of the derived identification error bounds in a robust controller design to obtain an exponentially stable closed loop.
    By means of an extensive numerical study, we showcase the interplay between the controller design and the derived identification error bounds. 
    Moreover, we note appealing connections of our results to indirect data-driven control of general nonlinear systems through Koopman operator theory and discuss how our results may be applied in this setup.
    \vspace*{-1.5\baselineskip}
\end{abstract}


\end{frontmatter}

%
%
\section{Introduction}
\vspace*{-\baselineskip}
Bilinear systems are an important class of nonlinear systems that naturally appears across different domains such as biological processes~\citep{Mohler1980}, socioeconomics~\citep{Mohler1973} but also in engineering, e.g., nuclear reactor dynamics~\citep{Mohler1973} and thermal control processes such as building control~\citep{underwood:2002}.
Further, the class of bilinear systems has recently received great attention for its ability to represent nonlinear systems through a higher-dimensional lifting, 
e.g., Carleman linearization or Koopman operator theory~\citep{mauroy:mezic:susuki:2020,surana:2016,huang:ma:vaidya:2018,strasser:worthmann:mezic:berberich:schaller:allgower:2026}.
Due to the wide-ranging occurrences of bilinear systems, there is significant interest in learning the behavior of a bilinear system from data.
However, currently there are only very few methods that allow to analyze the identification error from a finite-sample perspective. 
Such finite-sample results are of particular importance when it comes to indirect data-driven control of bilinear systems, where the identified system model is used to control the real system. 
Since usually only finite data can be collected and this data is often affected by noise, it is important to account for the introduced uncertainty to obtain end-to-end guarantees.
\par
\vspace*{-0.75\baselineskip}
%
%
%
\textbf{Related works~}
There exists a rich literature in classical system identification for both linear and nonlinear systems~\citep{ljung1998system}. 
The special case of bilinear systems has received considerable interest since many of the techniques used in linear system identification can be carried over to the bilinear setting~\citep{FNAIECH1987}. 
In particular,~\citet{Favoreel1999} generalize linear subspace identification to the bilinear setting under the assumption of white noise excitation. 
Further,~\citet{Hizir2012} reduce the bilinear identification problem to the identification of an equivalent linear model by choosing suitable sinusoidal inputs.
The problem of persistency of excitation and input selection for the identification of bilinear systems has been considered, e.g., by~\citet{Dasgupta1989} and~\citet{Sontag2009}.
\par 
\vspace*{-0.75\baselineskip}
Note that the previously discussed classical system identification literature only provides asymptotic results in the presence of stochastic noise, i.e., results that consider the case where the number of data collected goes to infinity.
Building on recent advances in high dimensional statistics~\citep{wainwright2019high,AbbasiYadkori2011}, first finite-sample system identification results have recently emerged for \ac{LTI} and certain classes of nonlinear systems.
For \ac{LTI} systems, where the \ac{OLS} estimator is predominantly used,~\citet{dean2020sample} provide individual identification error bounds for the unknown system matrices assuming that the available data is independent. 
Correlation in trajectory data is handled by~\citet{simchowitz2018learning} using the block martingale small-ball condition.
Allowing for dependent data comes at the cost of being restricted to marginally stable systems and not recovering individual identification error bounds on the unknown matrices. 
While the stability assumption is overcome in the works of~\citet{ShiraniFaradonbeh2018} and~\citet{sarkar2019near}, finding individual error bounds for each of the matrices from trajectory data is still an open problem. 
Extending the \ac{LTI} literature,~\citet{Foster2020} and~\citet{Sattar2022} provide a finite-sample identification analysis for generalized linear systems with a known nonlinearity.
When it comes to bilinear systems,~\citet{sattar2022finite} establish finite-sample identification error bounds for data collected from a single trajectory. 
However, their derived bound relies on a potentially restrictive stability assumption and comes in the form of a single upper bound of the identification errors for all the identified system matrices.
\citet{Sattar2025} extend previous works on bilinear systems by providing a finite sample identification analysis for partially observed bilinear systems.
Since we can only provide a brief overview of the field of non-asymptotic system identification, we refer to~\citet{Tsiamis2023} and~\citet{Ziemann2023} for more detailed discussions.
\par
\vspace*{-\baselineskip}
Despite the interest in applying statistical learning theory tools to bound identification errors, there have been comparably less works using the finite-sample error bounds for a robust controller design.
One important reason for this is that the bounds are often not directly usable for a (robust) controller design, and therefore providing end-to-end guarantees for an indirect data-driven control scheme may be difficult.
For the linear-quadratic regulator,~\citet{dean2020sample} establish an indirect data-driven control scheme with end-to-end guarantees. In the work of~\citet{mania2019certainty} this analysis is improved. Further,~\citet{Tsiamis22a} provide upper and lower bounds on the sample complexity of stabilizing \ac{LTI} systems using indirect data-driven control. 
\par
\vspace*{-\baselineskip}
While these developments provide valuable insights into the theoretical limits of \emph{linear} indirect data-driven control, they do not address corresponding extensions to bilinear systems. For this class of systems, there
exists a rich literature on model-based controller designs, including Lyapunov-based methods~\citep{pedrycz:1980,derese:noldus:1980}, bang-bang control with linear switching policy~\citep{longchamp:1980}, quadratic state feedback~\citep{gutman:1981,gutman:1980}, nonlinear state feedback~\citep{benallou:mellichamp:seborg:1988}, constant feedback~\citep{luesink:nijmeijer:1989}, or schemes for passive bilinear systems~\citep{lin:byrnes:1994}. 
Moreover,~\citet{huang:jadbabaie:1999} propose to view the state of bilinear systems as a scheduling variable, which leads to a convex controller design using results for \mbox{(quasi-)linear} parameter-varying systems.
Another approach is to use \acp{LMI} to design controllers for bilinear systems in a local region, see, e.g.,~\citet{amato:cosentino:fiorillo:merola:2009} for a polytopic region and~\citet{khlebnikov:2018} for an ellipsoidal region, or~\citet{coutinho:desouza:dasilva:caldeira:prieur:2019} for input-delayed systems.
Relying on robust control techniques, closed-loop stability guarantees for bilinear systems are derived in~\citet{strasser:berberich:allgower:2023b} using an \ac{LMI}-based controller in a pre-defined region, while~\citet{strasser:berberich:allgower:2025} design a globally stabilizing controller based on \ac{SOS} optimization.
However, most of the available results require model knowledge or are restricted to noise-free systems.
\par
\vspace*{-0.75\baselineskip}
%
%
\textbf{Contribution~}
In this work, we consider the problem of identifying a bilinear system from noisy data to control the underlying system with end-to-end guarantees. 
Specifically, we leverage tools from statistical learning theory to enable robust control of bilinear systems using collected data.
First, we present novel finite-sample error bounds for identifying bilinear systems from finite i.\,i.\,d. data.
Here, we use the control input to solve a set of linear and affine identification problems in order to identify the bilinear system from data.
We note that the novel finite-sample analysis of affine identification problems might be of independent interest.
Since the corresponding \ac{OLS} solutions do no longer depend on purely random matrices we leverage properties of symmetric matrices to analyze the random part and the determinist parts of the corresponding matrices separately. 
Combining this with (anti-)concentration inequalities allows us to provide high-probability identification error bounds.
We not only present a priori identification error bounds revealing the structural dependencies on key problem parameters, but also data-dependent identification error bounds that prove to be less conservative. 
Compared to~\citet{sattar2022finite}, where finite-sample identification error bounds from trajectory data are provided, the identification error bounds derived in this work are structurally tailored to indirect data-driven control. 
This enables combining the identification error bound with robust control approaches for bilinear systems. 
More precisely, we provide an easy-to-use algorithm to derive an indirect data-driven controller along with closed-loop stability guarantees.
To the best of our knowledge, this is the first work providing such an end-to-end result for finite data affected by stochastic noise with possibly unbounded support in the case of bilinear systems.
Further, we show that the proposed results may be applicable beyond bilinear systems through the Koopman operator and note appealing connections to Koopman-based indirect data-driven control of more general classes of nonlinear systems.
Finally, we showcase the effectiveness of the results in several numerical investigations, where we demonstrate the interplay between the controller design and the derived error bounds. 
\par
\vspace*{-\baselineskip}
%
%
%
\textbf{Outline~}
This paper is structured as follows. 
Section~\ref{sec:problem-setup} introduces the problem setup including the considered bilinear systems. 
In Section~\ref{sec:non-asymp-bounds}, we derive finite-sample identification error bounds for bilinear systems.
Then, we use the obtained bounds for the design of indirect data-driven controllers guaranteeing closed-loop exponential stability of bilinear systems in Section~\ref{sec:controller-design}.
Finally, we illustrate the effectiveness of the derived identification error bounds in comparison to Monte Carlo simulations as well as in the controller design in Section~\ref{sec:numerics}, before concluding the paper in Section~\ref{sec:conclusion}.
\par 
\vspace*{-0.75\baselineskip}
%
%
%
\textbf{Notation~}
We denote the set of integers between $a$ and $b$ by $\mathbb{N}_{[a,b]} = [a,b]\cap \mathbb{N}$.
The unit sphere in $\mathbb{R}^{n}$ is denoted by $\sph^{n-1}$. 
For a positive scalar $c\in \R_{>0}$ we denote a sphere centered around the origin of $\R^n$ with radius $c$ by $c\sph^{n-1}$.
Given a matrix $A$, we denote the spectral norm by $\Vert A\Vert_2$. 
The operation $[a]_i$ extracts the $i$-th element of the vector $a$ or the $i$-th column when applied to a matrix.
We denote matrix blocks that can be inferred from symmetry by $\star$, i.e., we write $\Lambda^\top \Sigma \Lambda =[\star]^\top \Sigma \Lambda$. By $\otimes$ we denote the Kronecker product. 
Further, we write $X \sim \mathcal{N}({\mu, \Sigma})$ if the random vector $X\in \R^{n_x}$ is Gaussian distributed with mean $\mu$ and covariance $\Sigma$.
We write $ Y\sim \subG{\sigma^2}$ if the random variable $Y\in \R$ is zero-mean sub-Gaussian with variance proxy $\sigma^2$.
Moreover, we write $ X\sim \subGvec{\sigma^2}{n_x}$ if the random vector $X\in \R^{n_x}$ is zero-mean sub-Gaussian with variance proxy $\sigma^2$, that is if the one-dimensional marginals $\langle X, v\rangle$ are zero-mean sub-Gaussian random variables with variance proxy $\sigma^2$ for all $v \in \sph^{n_x-1}$.
Finally, $Y \sim U(a)$ and $Y \sim \subExp{\nu^2}{\alpha}$ denote a random variable $Y\in \R$ which is uniformly distributed on $[-a, a]$ and sub-exponential with parameters $(\nu^2, \alpha)$, respectively.
%
%
\vspace*{-0.8\baselineskip}
\section{Problem setup}\label{sec:problem-setup}
\vspace*{-\baselineskip}
We consider an unknown bilinear system of the form 
\begin{equation}\label{eq:BiLinSysGen}
    x_+ = A x + B_0 u + \sum\nolimits_{i=1}^{n_u} [u]_i A_i x + w,
\vspace*{-0.5\baselineskip}
\end{equation}
where $w \simiid \subGvec{\sigma_w^2}{n_x}$ is unknown process noise, $x, x_+,\in \R^{n_x}$ are the state vector at the current time step and the next time step, respectively, and $u \in \R^{n_u}$ is a control input.
Note that by defining $B_i \coloneqq A_i + A$, the system~\eqref{eq:BiLinSysGen} can be equivalently described by 
\begin{subequations}\label{eq:BiLinSys}
    \vspace*{-0.25\baselineskip}
    \begin{align}
        x_+ &= A x + B_0 u + \sum\nolimits_{i=1}^{n_u} [u]_i (B_i - A) x + w \\ 
        &= A x + B_0 u +  A_{ux} (u \otimes x) +w,
        \\[-1.5\baselineskip]\nonumber
    \end{align}
\end{subequations}
where $A_{ux} = \begin{bmatrix} B_1 - A &\cdots & B_{n_u} - A \end{bmatrix}$.
We consider an indirect data-driven control scheme which consists of two steps.
First we identify the unknown matrices ${A \in \R^{n_x\times n_x}}$, ${B_0 \in \R^{n_x\times n_u}}$, ${B_1, \ldots, B_{n_u} \in \R^{n_x\times n_x}}$ from data and characterize the uncertainty of the estimates. 
Second, we deploy a robust control scheme accounting for the identification error to obtain a data-driven controller with end-to-end guarantees. 
\par
\vspace*{-0.75\baselineskip}
The structure in~\eqref{eq:BiLinSys} can be leveraged to reduce the nonlinear identification problem of identifying the bilinear system~\eqref{eq:BiLinSys} to $n_u+1$ linear identification problems. 
To this end, we conduct $n_u+1$ experiments in which we choose the fixed control inputs $u^{(i, \ell)} \equiv u^{e_0} \coloneqq 0$ and $u^{(i, \ell)}\equiv u^{e_i} \coloneqq e_i$ for $i \in \mathbb{N}_{[1, n_u]}$, respectively, where $e_i$ are the elements of the canonical basis of $\mathbb{R}^{n_u}$ and $i$ is the index of the experiment.\footnote{
    To simplify notation we will drop the dependence on $i$ for all data vectors, e.g., we use $u^{(\ell)}$ in place of $u^{(i, \ell)}$. The experiment $i \in \mathbb{N}_{[0, n_u]}$ will be clear from the context.
}
This yields the $n_u+1$ system descriptions
\begin{subequations}\label{eq:BiLinSysID}
    \vspace*{-0.05\baselineskip}
    \begin{alignat}{2}
         & \mathcal{S}_0: \quad x_+ &  & =  A x + w ,                  
         \label{eq:BiLinSys1}\\
         & \mathcal{S}_i: \quad x_+ &  & =  A x + [B_0]_i + (B_i - A) x + w \notag                              \\
         &                              &  & = [B_0]_i + B_i x + w, \quad \forall i =1, \dots, n_u,
         \label{eq:BiLinSys2}
         \\[-1.05\baselineskip]\nonumber
    \end{alignat}
\end{subequations}
describing the behavior of the unknown bilinear system~\eqref{eq:BiLinSys} under the respective control inputs. 
In the following, we consider the problem of identifying $\mathcal{S}_0, \dots \mathcal{S}_{n_u}$ which, as shown previously, is equivalent to identifying the bilinear system~\eqref{eq:BiLinSys}.
\vspace*{-0.85\baselineskip}
\begin{rem}
    We choose the canonical basis $\{e_1,\dots,e_{n_u}\}$ as inputs for simplicity. Particularly, any other basis $\{v_1,\dots, v_{n_u}\}$ of $\R^{n_u}$ could be chosen in addition to the zero input. 
    Clearly, there exists an invertible matrix $H$ that maps between the two bases, i.e., $v_i= H e_i$ for all $i\in\mathbb{N}_{[1, n_u]}$. 
    Applying the input $u^{(\ell)} \equiv v_i$ to system~\eqref{eq:BiLinSys} yields
    \vspace*{-0.5\baselineskip}
    \begin{align*}
        x_+ &= Ax + B_0 v_i + A_{ux} (v_i \otimes x) \\
        &= Ax + B_0 H e_i + A_{ux} (H e_i \otimes x) \\
        &= A x + B_0 H e_i + A_{ux} (H e_i\otimes I_{n_x}) x.
        \\[-1.25\baselineskip]
    \end{align*}
    Defining $\tilde{B}_i = A_{ux} (H e_i\otimes I_{n_x}) + A$ and $\tilde{B}_0= B_0 H$ yields $
        x_+ = \tilde{B}_i x + [\tilde{B}_0]_i  
    $%
    , i.e., a structurally identical identification problem to~\eqref{eq:BiLinSys2}. 
    Understanding the effect of the selected basis on the accuracy of identification and subsequently on the data-driven control scheme is an interesting direction for future work.
\end{rem}
\vspace*{-0.75\baselineskip}
To solve the linear and affine identification problems~\eqref{eq:BiLinSysID}, we resort to the \ac{OLS} estimator to obtain finite sample guarantees.
To this end, we collect $T_i$ samples from $\mathcal{S}_i$ for each $i\in\mathbb{N}_{[0, n_u]}$, where the number of samples will be specified in our analysis.
Then, the data collected from each experiment $i \in \mathbb{N}_{[0, n_u]}$ amounts to $\big\{x_+^{(\ell)}, x^{(\ell)}, u^{(\ell)}\big\}_{\ell =1}^{T_i}$.
Identifying the autonomous \ac{LTI} system~\eqref{eq:BiLinSys1} from finite data has already been considered, see, e.g.,~\citet{matni2019tutorial} for a detailed analysis.
However, the finite-sample identification of the affine system~\eqref{eq:BiLinSys2} is yet unsolved and is the main technical challenge for obtaining finite sample identification error bounds tailored to control. 
More precisely, the robust controller design requires identification error bounds which are proportional to the state and input.
For this reason, we seek finite sample identification error bounds that hold for all the unknown matrices individually, hence named individual identification error bounds hereafter. 
Obtaining individual, a priori identification error bounds from correlated data is an open problem in literature even in the case of linear systems~\citep{Ziemann2023}. 
To address this, we restrict the sampling according to the following assumption. 
\vspace*{-0.85\baselineskip}
\begin{assum}\label{ass:sampling}
    For each of the realizations $\mathcal{S}_i$, $i\in\mathbb{N}_{[0,n_u]}$, there exists $\sigma_x >0$ such that the sampled data $\big\{x_+^{{(\ell)}}, x^{{(\ell)}} \big\}_{\ell = 1}^{T_i}$ satisfies $x^{(\ell)} \simiid\subGvec{\sigma_x^2}{n_x}$.
\end{assum}

\vspace*{-2\baselineskip}
Note that, although potentially restrictive in practical applications, independence of the data is key for the proposed individual bounds. 
However, assuming the same distribution for each $\mathcal{S}_i$ is without loss of generality and the distribution can be selected to meet specified requirements.
In practice, Assumption~\ref{ass:sampling} can be satisfied by collecting the data from multiple trajectories, similar to approaches proposed by~\citet{dean2020sample,matni2019tutorial}.
Specifically, to obtain samples $x^{(\ell)}$ that satisfy Assumption~\ref{ass:sampling}, we can repeatedly initialize trajectories at the origin and then apply an input $u \simiid \subGvec{\sigma_u^2}{n_u}$ to the plant \eqref{eq:BiLinSys} for one time step.
Repeating this procedure $T_i$-times results in states $x^{(\ell)}$ that satisfy Assumption~\ref{ass:sampling} and hence can be used as starting points for the experiments with systems~\eqref{eq:BiLinSysID}. 
Note that, while the variance proxy $\sigma_x^2$ resulting from this procedure can be influenced by $\sigma_u^2$, it also depends on unknown system parameters.
This variance proxy enters the a-priori identification error bounds in Section~\ref{sec:APrioriBounds}, making them potentially hard to use in practice. 
The observed dependence is fundamental and cannot be avoided if a-priori identification error bounds are desired and the data cannot be sampled i.i.d.\ (cf., e.g., the error bounds presented by~\citet{Tsiamis2023,dean2020sample,sattar2022finite}). 
When it comes to the presented data-dependent identification error bounds (Section~\ref{sec:data-bounds}), which are the ones used in practice for the robust controller design (Section~\ref{sec:controller-design-controller}), this variance proxy is implicitly captured in the observed data and as a result they do not depend on unknown quantities and can be evaluated in practice.
\vspace*{-0.75\baselineskip}
\begin{rem}
    Compared to the works of~\citet{dean2020sample,matni2019tutorial} which rely on Gaussian noise and sampling, the extension to sub-Gaussian sampling allows for more correlations inside the sampled state vector.
    As a consequence, the constants in the burn-in time conditions are larger, but this more general analysis paves the way for applications in Koopman-based control (Section~\ref{sec:koop}).
\end{rem} 
\vspace*{-0.75\baselineskip}
Algorithm~\ref{algo:IDSampling} summarizes the proposed identification procedure, where the deployed sampling scheme is determined depending on the desired error bounds. 
\begin{algorithm}[t]
    \caption{Proposed identification algorithm}
    \begin{algorithmic}
        \Require Sampling scheme
        \For{$i \in \mathbb{N}_{[0, n_u]}$}
            \State Choose input $u^{e_i}$
            \For{$\ell \le T_i$}
                \State Sample state $x^{(\ell)}$ according to sampling scheme
                \State Evaluate bilinear system with $x^{(\ell)}$ and $u^{e_i}$
            \EndFor
        \EndFor
        \State Compute \ac{OLS} estimates for~\eqref{eq:BiLinSysID}
    \end{algorithmic}
    \label{algo:IDSampling}
\end{algorithm}
%
%
\vspace*{-0.75\baselineskip}
\section{Finite sample identification error bounds}\label{sec:non-asymp-bounds}
\vspace*{-0.75\baselineskip}
Next, we present high probability finite sample identification error bounds for each of the unknown elements in~\eqref{eq:BiLinSysID} in order to identify the bilinear system~\eqref{eq:BiLinSys}.
In particular, we use the \ac{OLS} estimator to identify the true system parameters from data. 
To this end, we define
\begin{equation*}                                             
    \theta_i 
    \coloneqq \begin{bmatrix}
        B_i & [B_0]_i
    \end{bmatrix}     
    \quad \forall i \in \mathbb{N}_{[1,n_u]}
    ,\quad
    y^{(\ell)} 
    \coloneqq \begin{bmatrix}
        {x^{(\ell)}}^\top & 1
    \end{bmatrix}^\top
\vspace*{-0.2\baselineskip}
\end{equation*}
such that the \ac{OLS} estimator is given by  
\begin{subequations}\label{eq:LeastSq}
    \begin{align}
        \hspace*{-0.04\linewidth}
        \hat{A} & \in \argmin_{A} \sum\nolimits_{\ell=1}^{T_0} \Vert x_{+}^{(\ell)}  - A x^{(\ell)} \Vert_2^2,                           
        \label{eq:LestSq1}\\
        \hspace*{-0.04\linewidth}
        \hat{\theta}_i & \in  \argmin_{\theta_i} \sum\nolimits_{\ell=1}^{T_i} \Vert x_{+}^{(\ell)} - \theta_i y^{(\ell)} \Vert_2^2 \quad \forall i\in \mathbb{N}_{[1,n_u]}.\!\!
        \label{eq:LestSq2}
    \end{align}
\end{subequations}
Further, we introduce the normalized regressors
\begin{equation}
    \xi^{(\ell)} \coloneqq \tfrac{x^{(\ell)}}{\sigma_x}\simiid \subGvec{1}{n_x}, \quad \zeta^{(\ell)} \coloneqq \begin{bmatrix}
        {\xi^{(\ell)}}^\top & 1 
        \end{bmatrix}^\top
\vspace*{-0.25\baselineskip}
\end{equation}
and, with slight abuse of notation, define
\begin{alignat*}{2}
    X_i &\coloneqq \begin{bmatrix}
            \xi^{(1)} & \cdots & \xi^{(T_i)}
        \end{bmatrix}^\top\hspace{-3pt},
    &\,
    Y_i &\coloneqq\begin{bmatrix}
            \zeta^{(1)} & \cdots & \zeta^{(T_i)}
        \end{bmatrix}^\top\hspace{-3pt},
    \\
    X_i^+ &\coloneqq \begin{bmatrix}
        {x_+^{(1)} } & \cdots & {x_+^{(T_i)}}
    \end{bmatrix}^\top\hspace{-3pt},
    &\,
    W_i &\coloneqq \begin{bmatrix}
            w^{(1)}  &  \cdots & w^{(T_i)}
        \end{bmatrix}^\top\hspace{-3pt}.
\end{alignat*}
Then, the closed-form solutions to~\eqref{eq:LeastSq} read
\begin{subequations}\label{eq:LsSolFull}
    \begin{align}
        \hat{A}^\top &= \tfrac{1}{\sigma_x} (X_0^\top X_0)^{-1}(X_0^\top X_0^+),
        \label{eq:LsSol1}
        \\
        \hat{\theta_i}^\top &= \left[\begin{smallmatrix}
        \frac{1}{\sigma_x} & 0 \\ 0 & 1
        \end{smallmatrix}\right] (Y_i^\top Y_i)^{-1}(Y_i^\top X_i^+), \quad \forall i \in \mathbb{N}_{[1,n_u]}.
        \label{eq:LsSol2} 
    \end{align}
\end{subequations}
Defining the normalized empirical covariance matrices
\begin{equation}\label{eq:DefM_full}
    M_0 \coloneqq \sum\nolimits_{\ell=1}^{T_0} \xi^{(\ell)} {\xi^{(\ell)}}^\top, 
    \,
    M_i \coloneqq \left[\begin{smallmatrix} 
        \sum_{\ell=1}^{T_i} \xi^{(\ell)} {\xi^{(\ell)}}^\top & \sum_{\ell=1}^{T_i} \xi^{(\ell)} \\
        \sum_{\ell=1}^{T_i} {\xi^{(\ell)}}^\top & T_i
    \end{smallmatrix}\right]
\vspace*{-0.5\baselineskip}
\end{equation}
for all $i \in \mathbb{N}_{[1,n_u]}$ leads to the identification errors
\begin{subequations}
    \begin{align}
        (\hat{A} - A)^\top &= \tfrac{1}{\sigma_x} M_0^{-1} (X_0^\top W_0),
        \label{eq:LQErrorII} 
        \\
        (\hat{\theta}_i - \theta_i)^\top &= \left[\begin{smallmatrix}
            \frac{1}{\sigma_x} & 0 \\ 0 & 1
            \end{smallmatrix}\right] M_i^{-1} (Y_i^\top W_i),\, \, \forall i \in \mathbb{N}_{[1,n_u]}.
        \label{eq:LQErrorI}
        \\[-1.25\baselineskip]\nonumber
    \end{align}
\end{subequations}
The identification error~\eqref{eq:LQErrorII} has been previously analyzed for Gaussian noise, e.g., by~\citet{matni2019tutorial}. 
We extend the result to sub-Gaussian noise and sampling.
\vspace*{-0.75\baselineskip}
\begin{thm}\label{th:ID_A}
    Consider the autonomous system~\eqref{eq:BiLinSys1}. Fix a failure probability $\delta\in (0,1)$ and let the data $\big\{x_+^{{(\ell)}}, x^{{(\ell)}} \big\}_{\ell = 1}^{T_0}$ be collected according to Assumption~\ref{ass:sampling}. If 
    \begin{equation}
        T_0 \ge 128 \log(8 \cdot 9^{n_x}/\delta),
    \end{equation}
    then the identification error~\eqref{eq:LQErrorII} of the \ac{OLS} estimate~\eqref{eq:LsSol1} is bounded by
    \begin{equation}
        \Vert \hat{A} - A \Vert_2 \le \frac{\sigma_w}{\sigma_x} \frac{16\sqrt{T_0 \log(4\cdot 9^{n_x}/\delta)}}{T_0}  
    \end{equation} 
    with probability at least $1-\frac{\delta}{2}$. 
\end{thm}%
\vspace*{-1.5\baselineskip}
\begin{pf}%
    First, using sub-multiplicativity of the norm,~\eqref{eq:LQErrorII} yields
    \begin{equation*}
        \Vert \hat{A} - A \Vert_2 \le \frac{1}{\sigma_x} \frac{\Vert X_0^\top W_0 \Vert}{\lambda_\mathrm{min}(M_0)}.
    \end{equation*}
    Then, we apply Propositions~\ref{prop:noiseTermSubG} and~\ref{prop:lowerboundCov} (Appendix~\ref{app:technical_results}) with $\frac{\delta}{4}$ and $c=\frac{1}{4}$ to obtain the result.
\end{pf}%
\vspace*{-1.5\baselineskip}
In the following, we use tools from \citet{vershynin2010introduction} and \citet{wainwright2019high} to derive a priori and data-dependent upper bounds on the identification error in~\eqref{eq:LQErrorI}. 
%
%
\subsection{A priori identification error bounds}\label{sec:APrioriBounds}
\vspace*{-0.75\baselineskip}
First, we provide novel a priori identification error bounds, which reveal fundamental dependencies on key parameters, e.g., the problem size or the desired confidence for the affine identification problem~\eqref{eq:BiLinSys2}.
\vspace*{-0.75\baselineskip}
\begin{thm}\label{th:sampleCompelxitySubG}
    \begin{subequations}
    Consider the unknown system $\mathcal{S}_i$ as defined in \eqref{eq:BiLinSys2} for any $i\in\mathbb{N}_{[1,n_u]}$. Fix a failure probability $\delta\in (0,1)$ and let the data $\big\{x_+^{{(\ell)}}, x^{{(\ell)}} \big\}_{\ell = 1}^{T_i}$ be collected according to Assumption~\ref{ass:sampling}. If 
    \begin{equation*}
        T_i \ge 64(3+ 2\sqrt{2}) \log(8n_u 9^{n_x}/\delta),
    \end{equation*}
    then the identification error~\eqref{eq:LQErrorI} of the \ac{OLS} estimate~\eqref{eq:LsSol2} is bounded by
    \begin{align*}
        \Vert  (\hat{B}_i- B_i) \Vert_2      
        &\leq \frac{\sigma_w}{\sigma_x} \frac{\frac{4\sqrt{10}}{3} \sqrt{2 T_i \log(4 n_u 9^{n_x}/\delta)}}{T_i/2 - \frac43 \sqrt{2 T_i \log(4 n_u  9^{n_x}/\delta) }}, \\
        \Vert  (\hat{[B_0]}_i- [B_0]_i) \Vert_2 
        &\leq \sigma_w \frac{\frac{4\sqrt{10}}{3} \sqrt{2 T_i \log(4 n_u 9^{n_x}/\delta)}}{T_i/2 - \frac43 \sqrt{2 T_i \log(4 n_u 9^{n_x}/\delta) }}
    \end{align*}
    with probability at least $1-\frac{\delta}{2 n_u}$.
    \end{subequations}
\end{thm}
\vspace*{-2\baselineskip}
\begin{pf}
    While this proof is structured similar to the proof of \ac{LTI} finite sample identification results (see, e.g.,~\citet{matni2019tutorial}), there are some key differences owing to the difficulties introduced by the affine structure in~\eqref{eq:BiLinSys2}. 
    Importantly, the regressor $y^{(\ell)}$ is not purely random.
    Thus, the matrix $M_i$ defined in~\eqref{eq:DefM_full} is not a purely random matrix and, hence, cannot be handled using the existing arguments.
    Analyzing these partially random quantities will pose the main technical difficulty.
    \par
    \vspace*{-0.75\baselineskip}
    Since we are interested in individual error bounds of $B_i$ and $[B_0]_i$, we observe that
    \begin{align*}
        (\hat{B}_i- B_i)^\top 
        &= \begin{bmatrix}
            I_{n_x} & 0_{n_x\times 1}
        \end{bmatrix}
        (\hat{\theta}_i - \theta_i)^\top 
        \\
        (\hat{[B_0]}_i- [B_0]_i)^\top 
        &= \begin{bmatrix}
            0_{1\times n_x} & 1
        \end{bmatrix}
        (\hat{\theta}_i - \theta_i)^\top.
    \end{align*}
    Exploiting~\eqref{eq:LQErrorI} results in the individual error bounds
    \begin{subequations}\label{eq:individError}
        \begin{align}
            (\hat{B}_i - B_i)^\top 
            &= \begin{bmatrix}\frac{1}{\sigma_x}I_{n_x} & 0_{n_x \times 1} \end{bmatrix}   
            M_i^{-1}
            Y_i^\top
            W_i ,
            \label{eq:indivError1}
            \\
            ([\hat{B}_0]_i  - [B_0]_i)^\top 
            &= \begin{bmatrix} 0_{1 \times n_x} & 1\end{bmatrix}   
            M_i^{-1}
            Y_i^\top
            W_i.
        \end{align}
    \end{subequations}
    Now, we take the norm of~\eqref{eq:individError} and use submultiplicativity of the matrix norm to obtain 
    \begin{subequations}\label{eq:BoundsRaw}
        \begin{align}    
            \|\hat{B}_i- B_i\|_2        
            &\leq \frac{1}{\sigma_x}\frac{\left\|
                Y_i^\top
                W_i 
            \right\| _2}{\lambda_{\mathrm{min}}(M_i)},
            \\
            \|\hat{[B_0]}_i- [B_0]_i\|_2 
            &\leq \frac{\left\|
                Y_i^\top
                W_i 
            \right\|_2}{\lambda_{\mathrm{min}}(M_i)},
        \end{align}
    \end{subequations}
    where $\lambda_\mathrm{min}(M_i) $ denotes the smallest eigenvalue of the matrix $M_i$.
    We split the analysis of the terms on the right-hand side into the analysis of the smallest eigenvalue of $M_i$ and controlling the norm in the numerator.
    \vspace*{-0.75\baselineskip}

    \textbf{Controlling the smallest eigenvalue of $M_i$.~}
    Since $M_i$ is not a purely random matrix we need to deploy different tools than the ones presented in~\citet{matni2019tutorial} to control $\lambda_\mathrm{min}(M_i)$.
    In particular, we proceed in two steps.
    First, we show that we can express the smallest eigenvalue of the full matrix as the smallest eigenvalue of the block diagonal terms and an error term depending on the off-diagonal elements.
    Then, we use Hoeffding's inequality to show that the off-diagonal elements are small compared to the block-diagonal terms if we collect enough samples.
    For the subsequent analysis we apply the Courant-Fisher minimax theorem~\citep[Theorem 8.1.2]{golub2013matrix} and consider $v \in \R^{n_x+1}$ with $\Vert v \Vert = 1$, such that we obtain%
    \vspace*{-2.5\baselineskip}
    \small%
    \begin{align*}
        &\lambda_\mathrm{min}(M_i) \hspace{-1pt} = \hspace{-3pt}\min_{v \in \mathbb{S}^{n_x}} 
        [\star]^\top
        \hspace{-2pt}
    \left[\begin{smallmatrix} 
        \sum_{\ell=1}^{T_i} \xi^{(\ell)} {\xi^{(\ell)}}^\top & \sum_{\ell=1}^{T_i} \xi^{(\ell)} \\
        \sum_{\ell=1}^{T_i} {\xi^{(\ell)}}^\top & T_i
    \end{smallmatrix}\right]  
    \hspace{-2pt} 
        \left[\begin{smallmatrix}
            v_1 \\ v_2
        \end{smallmatrix}\right]    
        \\
        &= \min_{v \in \sph^{n_x}} v_1^\top \Big(\sum_{\ell=1}^{T_i} \xi^{(\ell)} {\xi^{(\ell)}}^\top\Big)  v_1 + 2 v_2 v_1^\top \sum_{\ell=1}^{T_i} \xi^{(\ell)} + v_2^2 T_i                                                       
        \\
        & \ge \min_{v \in \sph^{n_x}} \hspace{-3pt} v_1^\top \Big(\sum_{\ell=1}^{T_i} \xi^{(\ell)} {\xi^{(\ell)}}^\top \Big)  v_1  + v_2^2 T_i - 2 \vert v_2\vert \Big\vert v_1^\top \sum_{\ell=1}^{T_i} \xi^{(\ell)} \Big\vert.
        \\[-1.9\baselineskip]
    \end{align*}
    \normalsize%
    Note that we can combine the first two terms into an eigenvalue condition on a block-diagonal matrix, i.e.,
    \vspace*{-0.75\baselineskip}
    \begin{align}\label{eq:minEigRefomulation}
        \lambda_\mathrm{min} (M_i)
        \geq \lambda_\mathrm{min}&\left(
            \left[\begin{smallmatrix}
                \sum_{\ell=1}^{T_i} \xi^{(\ell)} {\xi^{(\ell)}}^\top  
                & 0 
                \\ 
                0 
                & T_i
            \end{smallmatrix}\right]
        \right) 
        \nonumber\\ 
        &- \max_{v\in \sph^{n_x}} 2  \vert v_2\vert \Big\vert v_1^\top \sum\nolimits_{\ell=1}^{T_i} \xi^{(\ell)} \Big\vert.
        \\[-1.65\baselineskip]\nonumber
    \end{align}
    More precisely,~\eqref{eq:minEigRefomulation} shows that the minimum eigenvalue of the block-diagonal matrix as well as the error term serve as a lower-bound on the true minimum eigenvalue.
    \par
    \vspace*{-0.75\baselineskip}
    We first  derive an upper bound on the term \linebreak$\max_{v\in \sph^{n_x}} 2  \vert v_2\vert \vert v_1^\top \sum_{\ell=1}^{T_i} \xi^{(\ell)} \vert$.
    To this end, note that $\Vert v \Vert_2 = 1$ which implies $\Vert v_1 \Vert_2^2 + \vert v_2 \vert^2 = 1$. 
    Defining 
    $
        \bar{v}_1 = \frac{1}{\sqrt{1-v_2^2}} v_1 
    $
    yields $\bar{v}_1 \in \sph^{n_x-1}$. 
    Hence, we rewrite%
    \small%
    \vspace*{-2.5\baselineskip}
    \begin{equation*}
        \max_{v\in \sph^{n_x}} 2  \vert v_2\vert \Big\vert v_1^\top \sum_{\ell=1}^{T_i} \xi^{(\ell)} \Big\vert 
        =\! \hspace{-5pt} \max_{\substack{v_2 \in [-1, 1] \\ \bar{v}_1 \in \sph^{n_x-1}}} \hspace{-4pt} 2 \vert v_2\vert \sqrt{1-v_2^2} \Big\vert \bar{v}_1^\top \sum_{\ell=1}^{T_i} \xi^{(\ell)} \Big\vert,
    \vspace*{-0.75\baselineskip}
    \end{equation*}
    \normalsize%
    where we can maximize over $\bar{v}_1$ and $v_2$ separately. Thus, we use Lemma~\ref{lemma:concetrSubGSum} (Appendix~\ref{app:technical_results}) with $\frac{\delta}{4n_u}$ to obtain 
    \vspace*{-1.1\baselineskip}
    \begin{align*}
        &\max_{v\in \sph^{n_x}} 2  \vert v_2\vert \Big\vert v_1^\top \sum\nolimits_{\ell=1}^{T_i} \xi^{(\ell)} \Big\vert
        \\ 
        &\le \max_{v_2\in [-1, 1]}\tfrac{4}{3} \vert v_2 \vert \sqrt{1-v_2^2}  \sqrt{2 T_i \log(4 n_u \cdot9^{n_x}/\delta)} 
        \\
        &= \tfrac{2}{3} \sqrt{2 T_i \log(4 n_u \cdot9^{n_x}/\delta)}
    \\[-1.75\baselineskip]
    \end{align*}
    with probability at least $1-\frac{\delta}{4 n_u}$. 
    Here, the last step follows by plugging in the maximizer
    \vspace*{-0.75\baselineskip}
    \begin{equation*}
        v_2^*= \argmax_{v_2\in [-1, 1]} \vert v_2\vert \sqrt{1-v_2^2} = \tfrac{1}{\sqrt{2}}.
    \vspace*{-0.75\baselineskip}
    \end{equation*}
    Now, we consider%
    \small%
    \vspace*{-2.5\baselineskip}
    \begin{equation*}
        \lambda_\mathrm{min}\left(\left[\begin{smallmatrix}
            \sum\limits_{\ell=1}^{T_i} \xi^{(\ell)} {\xi^{(\ell)}}^\top & 0 \\ 0 & T_i
        \end{smallmatrix}\right]\right) 
        = \hspace{-1pt}\min\left\{ \hspace{-2.5pt}\lambda_\mathrm{min}\Big(\sum_{\ell=1}^{T_i} \xi^{(\ell)} {\xi^{(\ell)}}^\top\Big), T_i\hspace{-2.5pt}\right\}.
    \vspace*{-0.75\baselineskip}
    \end{equation*}
    \normalsize%
    Then, we use Proposition~\ref{prop:lowerboundCov} (Appendix~\ref{app:technical_results}) to obtain 
    \vspace*{-0.75\baselineskip}
    \begin{equation*}
        \Prob\left[\lambda_\mathrm{min}\left(\sum\nolimits_{\ell=1}^{T_i} \xi^{(\ell)} {\xi^{(\ell)}}^\top\right)\ge T_i(1-2c)^2\right] \ge 1- \tfrac{\delta}{4 n_u},
    \vspace*{-0.75\baselineskip}
    \end{equation*}
    where $c\in (0, \tfrac12)$. 
    Thus,
    \vspace*{-0.75\baselineskip}
    \begin{equation}\label{eq:EVFinal}
        \lambda_\mathrm{min}\left(M_i\right)
        \ge T_i(1-2c)^2 - \tfrac{4}{3} \sqrt{2 T_i \log(4 n_u 9^{n_x}/\delta)}
    \end{equation}
    with probability at least $\big(1-\frac{\delta}{4 n_u }\big)\big(1-\frac{\delta}{4 n_u}\big) \ge 1-\frac{\delta}{2 n_u}$ if $T_i \ge  \frac{8}{c^2} \log(8 n_u 9^{n_x}/\delta)$.
    To ensure that \eqref{eq:EVFinal} yields a non-trivial bound (and the inverses in \eqref{eq:BoundsRaw} exist), we need to impose the additional condition 
    \vspace*{-0.75\baselineskip}
    \begin{equation*}
        T_i(1-2c)^2 - \tfrac{4}{3} \sqrt{2 T_i \log(8 n_u  9^{n_x}/\delta)} > 0 
    \vspace*{-0.75\baselineskip}
    \end{equation*}
    which is satisfied if $T_i > \frac{32}{9(1-2c)^4} \log(8 n_u  9^{n_x}/\delta)$.
    We select $c  = \frac{\sqrt{2}-1}{2\sqrt{2}}$, which yields 
    \vspace*{-0.75\baselineskip}
    \begin{equation}\label{eq:EVFinal2}
        \lambda_\mathrm{min}\left(M_i\right)
        \ge \tfrac{T_i}{2} - \tfrac{4}{3} \sqrt{2 T_i \log(4 n_u \cdot 9^{n_x}/\delta) }
    \vspace*{-0.75\baselineskip} 
    \end{equation}
    with probability at least $1-\frac{\delta}{2n_u}$ if
    \vspace*{-0.75\baselineskip}
    \begin{align*}
        T_i &> \max\big\{
            \tfrac{128}{9}, 
            64(3+ 2\sqrt{2})
        \big\} \log(8 n_u 9^{n_x}/\delta)
        \\
        &= 64(3+2\sqrt{2}) \log(8 n_u 9^{n_x}/\delta).
        \\[-1.5\baselineskip]\nonumber
    \end{align*}
    %
    %
    %
    \textbf{Controlling the noise term.~}
    To handle the numerator in~\eqref{eq:BoundsRaw}, consider that 
    \vspace*{-0.75\baselineskip}
    \begin{align*}
        &\big\|
            Y_i^\top
            W_i \big\|_2 
        = \sup_{\substack{u \in \sph^{n_x}\\v\in \sph^{n_x-1} }} \sum\nolimits_{\ell=1}^{T_i} \left(u^\top \left[\begin{smallmatrix}
            \xi^{(\ell)} \\ 1
        \end{smallmatrix}\right]\right) \left((w^{(\ell)})^\top v\right) 
        \\
        &=  \sup_{\substack{u \in \sph^{n_x} \\ v\in \sph^{n_x-1} }} \sum\nolimits_{\ell=1}^{T_i}  \left(u_1^\top \xi^{(\ell)} + u_2\right) \left((w^{(\ell)})^\top v\right)
        \\
        &= \sup_{\substack{u \in \sph^{n_x}\\v\in\sph^{n_x-1} }} \sum\nolimits_{\ell=1}^{T_i}(u_1^\top \xi^{(\ell)})\left((w^{(\ell)})^\top v\right) + u_2 \left((w^{(\ell)})^\top v\right).
        \\[-1.5\baselineskip]\nonumber
    \end{align*}
    Using similar techniques as in the analysis of~\eqref{eq:minEigRefomulation} and introducing 
    $
        \bar{u}_1 = \frac{1}{\sqrt{1-u_2^2}} u_1
    $
    leads to 
    \vspace*{-\baselineskip}
    \begin{align}
        &\left\Vert
            Y_i^\top
            W_i \right\Vert _2
        \nonumber\\ 
        &= \hspace{-3pt}\sup_{\substack{v \in \sph^{n_x-1} \\ \bar u_1 \in \sph^{n_x-1} \\ u_2 \in [-1, 1] }} \sum_{\ell=1}^{T_i}\sqrt{1-u_2^2}(\bar u_1^\top \xi^{(\ell)})\big((w^{(\ell)})^\top v\big) + u_2 \big((w^{(\ell)})^\top v\big) 
        \nonumber\\
        &\le \sup_{u_2 \in [-1, 1]}
        \Biggl[
            \sqrt{1-u_2^2} \sup_{\substack{v\in \sph^{n_x-1} \\ \bar u_1 \in \sph^{n_x-1}}} 
            \left[
                \sum _{\ell=1}^{T_i}(\bar u_1^\top \xi^{(\ell)})\left((w^{(\ell)})^\top v\right)
            \right] 
        \nonumber\\ &\qquad\qquad\qquad
            + \vert u_2 \vert  \sup_{v \in \sph^{n_x-1}} 
            \left| 
                \sum\nolimits_{\ell=1}^{T_i}  (w^{(\ell)})^\top v 
            \right|
        \Biggr].
        \label{eq:noiseterm2}
    \\[-1.85\baselineskip]\nonumber
    \end{align} 
    Observe that 
    \vspace*{-0.5\baselineskip}
    \begin{equation*}
        \sup_{\substack{v\in \sph^{n_x-1} \\ \bar u_1 \in \sph^{n_x-1}}} \left[
            \sum_{\ell=1}^{T_i}(\bar u_1^\top {\xi^{(\ell)}})\left((w^{(\ell)})^\top v\right) 
        \right] 
        = \left\| 
            \sum_{\ell = 1}^{T_i} \xi^{(\ell)} {w^{(\ell)} }^\top 
        \right\|_2,
    \vspace*{-0.75\baselineskip}
    \end{equation*}
    i.e., we can apply Proposition~\ref{prop:noiseTermSubG} (Appendix~\ref{app:technical_results}) with $\frac{\delta}{4n_u}$ to obtain that if $T_i \ge \frac12 \log(4n_u 9^{2n_x}/\delta)$, then
    \vspace*{-0.75\baselineskip}
    \begin{align}\label{eq:crossterm}
        \sup_{\substack{v\in \sph^{n_x-1} \\ \bar u_1 \in \sph^{n_x-1}}} \hspace{-3pt} \sum _{\ell=1}^{T_i}(\bar u_1^\top y^{(\ell)})\big((w^{(\ell)})^\top v\big) \hspace{-3pt} \le \hspace{-2pt} 4 \sigma_w \sqrt{T_i \log(4 n_u \nicefrac{9^{2n_x}}{\delta})}
        \nonumber\\[-1.25\baselineskip]\\[-1.5\baselineskip]\nonumber
    \end{align}
    holds with probability at least $1-\frac{\delta}{4n_u}$.
    Further, we use Lemma~\ref{lemma:concetrSubGSum} (Appendix~\ref{app:technical_results}) to deduce the upper bound 
    \vspace*{-0.75\baselineskip}
    \begin{equation}\label{eq:noiseConcetration}
        \hspace*{-0.04\linewidth}
        \sup_{v \in \sph^{n_x-1}} \left\vert v^\top \sum\nolimits_{\ell=1}^{T_i} w^{(\ell)} \right\vert \le \tfrac{4}{3} \sigma_w \sqrt{2T_i \log(4n_u 9^{n_x}/\delta)}
    \end{equation}
    with probability at least $1-\frac{\delta}{4n_u}$.
    Union bounding \eqref{eq:crossterm} and \eqref{eq:noiseConcetration} and plugging the result into \eqref{eq:noiseterm2} leads to
    \vspace*{-0.75\baselineskip}
    \begin{align*}
        &\left\| 
        Y_i^\top
        W_i \right\|_2 
        \le \sup_{u_2\in [-1, 1]} \sqrt{1-u_2^2}  4 \sigma_w \sqrt{T_i \log(4 n_u 9^{2n_x}/\delta)} 
        \\
        &\qquad \qquad\qquad\; + \vert u_2\vert \tfrac{4}{3} \sigma_w \sqrt{2T_i \log(4n_u 9^{n_x}/\delta)} 
        \\
        &\le \sup_{u_2\in [-1, 1]} \sqrt{1-u_2^2}  4 \sigma_w \sqrt{2 T_i \log(4 n_u 9^{n_x}/\delta)} 
        \\
        &\qquad+ \vert u_2\vert \tfrac{4}{3} \sigma_w \sqrt{2T_i \log(4n_u 9^{n_x}/\delta)} 
        \\
        &= \sup_{u_2\in [-1, 1]} 4 \sigma_w \sqrt{2 T_i \log(4 n_u 9^{n_x}/\delta)}\left( \sqrt{1-u_2^2} + \tfrac{\vert u_2\vert}{3} \right)
    \\[-1.75\baselineskip]
    \end{align*}
    with probability at least $1-\frac{\delta}{2n_u}$ if $T_i \ge \frac12 \log(4n_u 9^{2n_x}/\delta)$.
    Finally, plugging in the maximizer $u_2^* = \frac{1}{\sqrt{10}}$ yields 
    \vspace*{-0.75\baselineskip}
    \begin{equation}
        \left\| 
        Y_i^\top
        W_i \right\|_2 \le \tfrac{4\sqrt{10}}{3} \sigma_w \sqrt{2 T_i \log(4 n_u 9^{n_x}/\delta)}.
        \label{eq:noiseControlFinal}
    \vspace*{-0.5\baselineskip}
    \end{equation}
    %
    %
    %
    \textbf{Combining the previous results.~}
    Taking \eqref{eq:EVFinal2} and \eqref{eq:noiseControlFinal}, plugging them into \eqref{eq:BoundsRaw}, and using union bound arguments leads to the desired result. 
    Further, we compare the two burn-in time conditions 
    \vspace*{-0.75\baselineskip}
    \begin{align*}
        \max&\left\{\tfrac{1}{2} \log(4n_u 9^{2n_x}/\delta), 64(3+ 2\sqrt{2}) \log(8n_u 9^{n_x}/\delta)\right\} 
        \\
        &\quad= 64(3+ 2\sqrt{2}) \log(8n_u 9^{n_x}/\delta) \quad \forall n_x, n_u \ge 1
    \\[-1.5\baselineskip]
    \end{align*}
    which concludes the proof.
\end{pf}
\vspace*{-1.5\baselineskip}
Note that like in the \ac{LTI} case assuming i.i.d.\ data allows us to provide error bounds on the individual matrices independently of the stability properties of the system considered. 
Further, the bounds in Theorem~\ref{th:sampleCompelxitySubG} can be computed \emph{before} collecting data, relying only on knowledge of the system dimensions and the noise variance.
%
%
\vspace*{-0.5\baselineskip}
\subsection{Data-dependent identification error bounds}\label{sec:data-bounds}
\vspace*{-0.5\baselineskip}
Depending on the application it might not be necessary to provide a priori identification error bounds, i.e., before data collection. 
Instead, one can turn to data-dependent error bounds that account only for the data observed and, hence, are less conservative.
To this end, recall the matrices $M_i$ defined in \eqref{eq:DefM_full} and note that they can be evaluated from data, i.e., we do not need to control the respective inverses. 
Thus, we can leverage the derivations from the previous section to obtain individual bounds for each of the unknown matrices.
\vspace*{-0.5\baselineskip}
\begin{cor}\label{co:ThwithMFromData}
    Consider the unknown system $\mathcal{S}_i$ as defined in \eqref{eq:BiLinSys2} for any $i\in\mathbb{N}_{[1,n_u]}$. Fix a failure probability $\delta\in (0,1)$ and let the data $\{x^{(\ell)}_+, x^{(\ell)}\}_{\ell=1}^{T_i}$ be collected according to Assumption~\ref{ass:sampling}.
    If $T_i \ge \frac12 \log(2 n_u 9^{2n_x}/\delta)$, then the identification error of the \ac{OLS} estimate~\eqref{eq:LsSol2} satisfies
    \begin{subequations}
        \begin{align*}
            \Vert  (\hat{B}_i- B_i) \Vert_{2}        
            & \le \frac{\sigma_w}{\sigma_x} \frac{\frac{4\sqrt{10}}{3}  \sqrt{2 T_i \log(2 n_u\cdot 9^{n_x}/\delta)}}{\lambda_{\mathrm{min}}(M_i)}, \\
            \Vert  (\hat{[B_0]}_i- [B_0]_i) \Vert_2 
            & \le \sigma_w \frac{\frac{4\sqrt{10}}{3} \sqrt{2 T_i \log(2 n_u\cdot 9^{n_x}/\delta)}}{\lambda_{\mathrm{min}}(M_i)}
        \end{align*} 
    \end{subequations}
    with probability at least $1-\frac{\delta}{2 n_u}$, where the matrix $M_i$ is defined in~\eqref{eq:DefM_full}. 
    If $M_i$ has zero as an eigenvalue, we define the inverse of that eigenvalue to be infinity.
\end{cor}
\vspace*{-0.75\baselineskip}
Alternatively, we can use similar proof techniques to \citet[Proposition 3]{dean2020sample} to obtain ellipsoidal, data-dependent identification error bounds.
\vspace*{-0.75\baselineskip}
\begin{lem}\label{co:data-bounds-ellipsoidal}
    Consider the unknown system $\mathcal{S}_i$ defined in \eqref{eq:BiLinSys2} for any $i\in\mathbb{N}_{[1,n_u]}$ and with $w^{(\ell)} \stackrel{\text{i.i.d.}}{\sim} \mathcal{N}(0, \sigma_w^2 I_{n_x})$. Fix a failure probability $\delta\in (0,1)$ and let the data $\{x^{(\ell)}_+, x^{(\ell)}\}_{t=1}^{T_i}$ be sampled i.i.d.\ with $T_i \ge n_x + 1$.
    Define
    \vspace*{-0.5\baselineskip}
    \begin{equation*}
        C_1(n_x, \delta) = \sigma_w^2\left(\sqrt{n_x+1}+ \sqrt{n_x}+ \sqrt{2\log({2 n_u}/{\delta})}\right)^2.
    \vspace*{-0.5\baselineskip}
    \end{equation*}
    Then, with probability at least $1-\frac{\delta}{2n_u}$, we have 
    \vspace*{-0.65\baselineskip}
    \begin{multline*}
        \begin{bmatrix}
            \star
        \end{bmatrix}
        \begin{bmatrix}
            (\hat{B}_i - B_i)^\top \\ 
            ([\hat{B}_0]_i - [B_0]_i)^\top 
        \end{bmatrix}^\top 
        \\ 
        \preceq C_1(n_x, \delta)
        \begin{bmatrix} 
        \sum_{\ell=1}^{T_i} x^{(\ell)} {x^{(\ell)}}^\top & \sum_{\ell=1}^{T_i} x^{(\ell)} \\
        \sum_{\ell=1}^{T_i} {x^{(\ell)}}^\top & T_i
        \end{bmatrix}^{-1}
        .
        \vspace*{-0.8\baselineskip}
    \end{multline*}
    If the empirical covariance matrix has zero as an eigenvalue, we define the inverse of that eigenvalue to be infinity.
\end{lem}
\vspace*{-1.75\baselineskip}
\begin{pf}
    We only provide a short version of the proof, for a more detailed version we refer to the proof of~\citet[Proposition V.1]{matni2019tutorial}. First, define $
        \tilde Y_i^\top = \left[\begin{smallmatrix}
            \sigma_x & 0 \\ 0 & 1
        \end{smallmatrix}\right] Y_i^\top
    $ and $
        E^\top =\begin{bmatrix}
            (\hat{B}_i - B_i) &  
            ([\hat{B}_0]_i - [B_0]_i)
        \end{bmatrix}
    $.
    Assuming $T_i \ge n_x +1$, the singular value decomposition of $\tilde Y_i$ is given by $\tilde Y_i = U \Lambda V^\top$. 
    Hence, if the inverse of $\Lambda$ exists, we obtain 
    \begin{align*}
        E E^\top &= V \Lambda^{-1} U^\top W_i W_i^\top U \Lambda^{-1}V^\top \preceq \Vert U^\top W\Vert_2^2 (\tilde Y_i^\top \tilde Y_i)^{-1}.
    \end{align*}  
    Each element of the matrix $U^\top W_i \in \R^{(n_x +1) \times n_x}$ is i.i.d.\ $\N(0, \sigma_w^2) $ and, thus, we apply~\citet[Corollary 5.35]{vershynin2010introduction} to show $\Vert U^\top W_i\Vert_2 \le \sigma_w \big(\sqrt{n_x +1} + \sqrt{n_x} + \sqrt{2 \log({2n_u}/{\delta})}\big)$ with probability at least $1-\frac{\delta}{2n_u}$.
\end{pf}
\vspace*{-1.65\baselineskip}
Note that, Lemma~\ref{co:data-bounds-ellipsoidal} only requires the data to be sampled i.i.d.\ but does not pose any additional requirements on the sampling distribution. This provides same additional flexibility that can be leveraged in practice.
Lemma~\ref{co:data-bounds-ellipsoidal} can be straightforwardly 
extended to sub-Gaussian noise using~\citet[Proposition 5.39]{vershynin2010introduction} instead of~\citet[Corollary 5.35]{vershynin2010introduction}.
%
%
\vspace*{-0.75\baselineskip}
\subsection{Sample complexity of identifying bilinear systems}
\vspace*{-0.75\baselineskip}
To obtain finite sample identification error bounds for the bilinear system~\eqref{eq:BiLinSys}, we apply Algorithm~\ref{algo:IDSampling} and combine the results from Sections~\ref{sec:APrioriBounds} and~\ref{sec:data-bounds} using union bound arguments.
\vspace*{-0.85\baselineskip}
\begin{thm}\label{th:ApproxAll}
    Consider Algorithm~\ref{algo:IDSampling} with data collected from the bilinear system~\eqref{eq:BiLinSys} according to Assumption~\ref{ass:sampling}. 
    Fix a failure probability $\delta \in (0,1)$.
    If 
    \begin{equation*}
        T_0 \geq \bar T_0
        \quad\text{and}\quad
        T_i \geq \bar T_i   \quad \forall i \in\mathbb{N}_{[1,n_u]},
    \end{equation*} 
    then Algorithm~\ref{algo:IDSampling} results in estimates $\hat{A}$, $\hat{B}_0$, $\hat{B}_1$, $\dots$, $\hat{B}_{n_u}$ that satisfy 
    \begin{subequations}
        \label{eq:errorBoundIndiv}
        \begin{alignat}{2}
            \Vert \hat{A} - A\Vert_2 &\le \varepsilon_A,&& \\
            \Vert \hat{B}_i - B_i \Vert_2 &\le \varepsilon_{B_i} \quad &&\forall i \in\mathbb{N}_{[1,n_u]}, \\
            \Vert [\hat{B}_0]_i - [B_0]_i \Vert_2 &\le \varepsilon_{[B_0]_i}  \quad &&\forall i \in\mathbb{N}_{[1,n_u]}
        \end{alignat}
    \end{subequations}
    with probability at least $1-\delta$, where identification error bounds and burn-in times are specified as follows:
    \vspace*{-0.75\baselineskip}
    \begin{enumerate}
        \item A priori identification error bounds:
        \vspace*{-0.1\baselineskip}
        \begin{subequations}
            \label{eq:prioriBoundsAll}
            \begin{align}
                \varepsilon_A 
                &= \frac{\sigma_w}{\sigma_x} \frac{16\sqrt{T_0 \log(4\cdot {9^{n_x}}/{\delta})}}{T_0},
                \label{eq:prioriErrorA} \\ 
                \varepsilon_{B_i} 
                &= \frac{\sigma_w}{\sigma_x} \frac{\frac{4\sqrt{10}}{3} \sqrt{2 T_i \log(4 n_u 9^{n_x}/\delta)}}{T_i/2 - \frac43 \sqrt{2 T_i \log(4 n_u  9^{n_x}/\delta) }}, 
                \label{eq:prioriErrorBi} \\
                \varepsilon_{[B_0]_i} 
                &= \sigma_w \frac{\frac{4\sqrt{10}}{3} \sqrt{2 T_i \log(4 n_u 9^{n_x}/\delta)}}{T_i/2 - \frac43 \sqrt{2 T_i \log(4 n_u 9^{n_x}/\delta) }},
                \label{eq:prioriErrorB0} \\
                \bar T_0 
                &= 128 \log(8 \cdot {9^{n_x}}/{\delta}), 
                \label{eq:prioriSampleA}\\
                \bar T_i 
                &= 64(3+ 2\sqrt{2}) \log(8n_u 9^{n_x}/\delta).
                \label{eq:prioriSampleBi}
                \\[-\baselineskip]\nonumber
            \end{align}
        \end{subequations}
        \item Data-dependent identification error bounds:
        \vspace*{-0.1\baselineskip}
        \begin{subequations}
            \label{eq:dataBoundsAll}
            \begin{align}
                \varepsilon_A 
                &= \frac{\sigma_w}{\sigma_x} \frac{4\sqrt{T_0 \log(4\cdot {9^{n_x}}/{\delta})}}{\lambda_{\mathrm{min}}(M_0)},
                \label{eq:dataErrorA}\\
                \varepsilon_{B_i} 
                &=  \frac{\sigma_w}{\sigma_x} \frac{\frac{4\sqrt{10}}{3}  \sqrt{2 T_i \log(2 n_u\cdot 9^{n_x}/\delta)}}{\lambda_{\mathrm{min}}(M_i)}, 
                \label{eq:dataErrorBi} \\
                \varepsilon_{[B_0]_i} 
                &= \sigma_w \frac{\frac{4\sqrt{10}}{3} \sqrt{2 T_i \log(2 n_u 9^{n_x}/\delta)}}{\lambda_{\mathrm{min}}(M_i)},
                \label{eq:dataErrorB0}\\
                \bar T_0 
                &= \tfrac{1}{2} \log(2 \cdot 9^{n_x}/{\delta}),
                \label{eq:dataSampleA}\\
                \bar T_i 
                &= \tfrac{1}{2} \log(2 n_u 9^{2n_x}/\delta).
                \label{eq:dataSampleBi}
            \end{align}
        \end{subequations}
    \end{enumerate}
\end{thm}
\vspace*{-1.75\baselineskip}
\begin{pf}
    This result follows directly by using Theorem~\ref{th:sampleCompelxitySubG} (Corollary~\ref{co:ThwithMFromData}) and (the data-dependent version of) Theorem~\ref{th:ID_A} and leveraging union bound arguments. 
\end{pf}
\vspace*{-1.65\baselineskip}
While we observe in Section~\ref{sec:numerics-error-bounds} that the a priori identification error bounds~\eqref{eq:prioriBoundsAll} are less tight than the data-dependent identification ones~\eqref{eq:dataBoundsAll}, they provide the possibility to bound the amount of uncertainty in the estimates before running the experiment.
Further, the a priori identification error bounds provide additional insights that help to understand the difficulty in identifying the system matrices in terms of the sample complexity. 
First, we can observe that the identification of every unknown matrix scales with $\bigO(\nicefrac{1}{\sqrt{T}})$ which is the known rate of \ac{OLS} for linear systems. 
Further, the problem size influences the identification errors~\eqref{eq:prioriErrorA}-\eqref{eq:prioriErrorB0} of order $\bigO(\sqrt{n_x \log(n_u)})$, whereas the overall number of samples $T = T_0 + n_u T_i$ needs to be of order $\bigO(n_x(n_u+1))$. 
Equivalent to the linear case, the failure probability enters inversely inside $\log$-terms and lastly, we observe the \ac{SNR} $\sigma_x/\sigma_w$ for the identification error bounds of $A$ and $B_i$, $\forall i \in \mathbb{N}_{[1,n_u]}$, whereas the identification error of $[B_0]_i$ lacks the dependence on $\sigma_x$. 
This is because $[B_0]_i$ enters affinely in~\eqref{eq:BiLinSys2}, i.e., the sampling variance cannot influence the rate of identifying $[B_0]_i$.
\par 
\vspace*{-\baselineskip}
Equivalently, we can use the ellipsoidal, data-dependent identification error bounds to obtain the following result. 
\vspace*{-1.75\baselineskip}
\begin{thm}\label{th:ApproxAllData}
    Consider Algorithm~\ref{algo:IDSampling} with i.i.d.\ data collected from the bilinear system~\eqref{eq:BiLinSys}. 
    Fix a failure probability $\delta \in (0,1)$ and let $w^{(\ell)} \simiid\N(0, \sigma_w^2I_{n_x})$. 
    If 
    \begin{equation*}
        T_0 \ge n_x \quad \text{and} \quad T_i \ge n_x + 1 \quad \forall i \in \mathbb{N}_{[1,n_u]},
    \vspace*{-0.5\baselineskip}
    \end{equation*} 
    then Algorithm~\ref{algo:IDSampling} results in estimates $\hat{A}$, $\hat{B}_0$, $\hat{B}_1$, $\dots$, $\hat{B}_{n_u}$ that satisfy
    \begin{subequations}\label{eq:ellipBoundsTotal}
        \begin{align}
           (\hat{A} - A)^\top (\hat{A} - A) &\preceq \mathcal{E}_{A}, 
           \\
            \begin{bmatrix}
                \star 
            \end{bmatrix}
            \begin{bmatrix}
                (\hat{B}_i - B_i)^\top \\ 
                ([\hat{B}_0]_i - [B_0]_i)^\top 
            \end{bmatrix}^\top &\preceq \mathcal{E}_{B_i} \quad \forall i \in \mathbb{N}_{[1,n_u]}
            \label{eq:ellipBoundsTotalB}
        \end{align} 
        with probability at least $1-\delta$, where 
        \vspace*{-0.5\baselineskip}
        \begin{align*}
            \mathcal{E}_{A}&= 
            \sigma_w^2
            \left(2\sqrt{n_x}+ \sqrt{2\log\left({2}/{\delta}\right)}\right)^2
            \left(\sum\nolimits_{\ell=1}^{T_0} x^{(\ell)} {x^{(\ell)}}^\top\right)^{-1}
            \\
            \mathcal{E}_{B_i} &= \sigma_w^2\left(\sqrt{n_x+1}+ \sqrt{n_x}+ \sqrt{2\log\left({2 n_u}/{\delta}\right)}\right)^2 \\ & \qquad \cdot
            \begin{bmatrix} 
                \sum_{\ell=1}^{T_i} x^{(\ell)} {x^{(\ell)}}^\top & \sum_{\ell=1}^{T_i} x^{(\ell)} \\
                \sum_{\ell=1}^{T_i} {x^{(\ell)}}^\top & T_i
            \end{bmatrix}^{-1}.
        \end{align*}
        If the empirical covariance matrices have zero as an eigenvalue, we define the inverse of that eigenvalue to be infinity.
    \end{subequations}
\end{thm}
\vspace*{-2\baselineskip}
\begin{pf}
    The result directly follows by using Lemma~\ref{co:data-bounds-ellipsoidal} and \citet[Proposition 2.4]{dean2020sample} with $\frac{\delta}2$ followed by union bound arguments. 
\end{pf}
\vspace*{-2\baselineskip}
With this, we have established a priori and data-dependent finite sample identification error bounds for the identification of bilinear systems from i.i.d.\ data.
While the results hold for bilinear systems, Koopman operator theory provides an appealing tool to extend the results of this work to more general nonlinear systems. While a detailed analysis is out of the scope of this work, we sketch some of the links between our results and the general nonlinear case in the following.
%
%
\vspace*{-0.85\baselineskip}
\subsection{Implications for data-driven control of nonlinear systems}\label{sec:koop}
\vspace*{-0.85\baselineskip}
As already discussed, Koopman operator theory~\citep{koopman:1931,mauroy:mezic:susuki:2020} allows to accurately represent nonlinear systems by higher-dimensional bilinear systems~\citep{surana:2016,huang:ma:vaidya:2018}.
Identifying this lifted bilinear system from data collected from the true system is an active field of research.
Although (extended) dynamic mode decomposition~\citep{williams:kevrekidis:rowley:2015} is shown to suitably approximate the Koopman operator~\citep{korda:mezic:2018b,bevanda:sosnowski:hirche:2021} while being scalable to large-scale systems and robust w.r.t. noise~\citep{bevanda:driessen:iacob:toth:sosnowski:hirche:2024,meanti2024estimating}, finite sample identification error bounds are usually hard to obtain for noisy systems~\citep{mezic2022numerical,nuske:peitz:philipp:schaller:worthmann:2023,philipp2024extended}.
One particular challenge lies in the fact that we cannot sample from the high-dimensional lifted state-space directly, but only from the lower-dimensional original state-space, where the two are related by known lifting functions.
Here, we emphasize that sampling in the original state space and lifting the samples afterwards does not violate the assumptions of Theorem~\ref{th:ApproxAllData}. 
Thus, this result can still be applied to a setting where the system is bilinear in a lifted state space. 
Regarding Theorem~\ref{th:ApproxAll}, the following proposition demonstrates for a particular choice of a lifting function, which has been widely used in the Koopman literature~\citep{mauroy:mezic:susuki:2020}, that sub-Gaussian sampling in the lifted state space (Assumption~\ref{ass:sampling}) can still be satisfied, enabling the application of Theorem~\ref{th:ApproxAll} in this setup. 
\vspace*{-1.25\baselineskip}
\begin{lem}\label{lemma:distOflifting}
    Consider a scalar random variable $x\stackrel{\text{i.i.d.}}{\sim} \uniform{a}$. Then, the random vector $\xi = \begin{bmatrix}
        x & \sin(x)
    \end{bmatrix}^\top$
    is sub-Gaussian distributed with variance proxy
    \vspace*{-0.25\baselineskip}
    \begin{equation*}
        \sigma^2 \le \begin{cases}
            2a +1, & \text{if } a \in (0, 1], \\
            a^2+2a, & \text{if } a \in (1, \infty).
        \end{cases}
    \end{equation*}
\end{lem}
\vspace*{-2.5\baselineskip}
\begin{pf} 
    See Appendix~\ref{app:proofdistOfLifting}.
\end{pf}
\vspace*{-2\baselineskip}
Although we provide the proof for a scalar variable $x$ for clarity of exposition, Lemma~\ref{lemma:distOflifting} can be easily extended to vector-valued random variables, where $\sin(\cdot)$ is applied element-wise.
While we show that a suitable sampling in the original state space ensures sub-Gaussian sampling in the lifted space for a specific lifting function, this remains an open question for general lifting functions.
However, we conjecture that Lemma~\ref{lemma:distOflifting} can be extended to other classes of lifting functions using bounded sampling, where similar results hold for Lipschitz continuous lifting functions~\citep[Theorem 2.26]{wainwright2019high}. 
Further, note that the derived upper bound of the variance proxy is not sharp, but shows that lifted samples are sub-Gaussian distributed.
In particular, the identification error bounds in Theorem~\ref{th:ApproxAll} require the exact variance proxy $\sigma_x^2$ or a lower bound which would result in a underestimation of the true \ac{SNR}.
Finding the exact variance proxy (or a tight lower bound) and a dedicated analysis for commonly used lifting functions are interesting directions for future research.
%
%
\vspace*{-\baselineskip}
\section{Controller design for bilinear systems}\label{sec:controller-design}
\vspace*{-0.75\baselineskip}
For the controller design, we consider the system representation in~\eqref{eq:BiLinSys} but focus on the noise-free part of the dynamics, i.e., our control objective is nominal stabilization.
In particular, we assume that the only uncertainty in the system dynamics arises from the identification error, which is common in the literature on (stochastic) data-driven control; see, e.g.,~\citet{waarde:camlibel:mesbahi:2020,martin:schon:allgower:2023b,faulwasser:ou:pan:schmitz:worthmann} and the references therein.
More precisely, we express~\eqref{eq:BiLinSys} in terms of the \ac{OLS} estimates $\hat{A}$, $\hat{B}_0=\begin{bmatrix}
    [\hat{B}_0]_1 & \cdots & [\hat{B}_0]_{n_u}
\end{bmatrix}$, $\hat{B}_1,\ldots,\hat{B}_{n_u}$ and define $\hat{A}_{ux} \coloneqq \begin{bmatrix}
    \hat{B}_1 - \hat{A} & \cdots & \hat{B}_{n_u} - \hat{A}
\end{bmatrix}$ to obtain the uncertain bilinear system
\vspace*{-0.1\baselineskip}
\begin{equation}\label{eq:BilinSysIdentified}
    x_t^+
    = \hat{A} x_t + \hat{B}_0 u_t + \hat{A}_{ux} (u_t \otimes x_t) + r(x_t,u_t).
\vspace*{-0.1\baselineskip}
\end{equation}
Here, $r(x,u)$ is the residual capturing the identification error resulting from the \ac{OLS} estimation and is given by
\begin{equation}\label{eq:residual-sampling-error}
    r(x,u) \coloneqq (A-\hat{A})x + (B_0-\hat{B}_0)u + (A_{ux} - \hat{A}_{ux}) (u\otimes x).
\vspace*{-0.1\baselineskip}
\end{equation}
In the following, we demonstrate that the non-asymptotic identification error bounds derived in Section~\ref{sec:non-asymp-bounds} are suitable for robust control of bilinear systems. Therefore, we follow the design proposed in~\citet{strasser:berberich:allgower:2023b,strasser:schaller:worthmann:berberich:allgower:2024b} in the context of a Koopman-based bilinear surrogate model. 
To this end, we bound the identification error by a quadratic function of state and input, i.e.,
\vspace*{-0.75\baselineskip}
\begin{equation}\label{eq:proportional-bound-residual}
    \|r(x,u)\|_2^2 
    \leq 
    \begin{bmatrix} x \\ u \end{bmatrix}^\top 
    Q_\Delta 
    \begin{bmatrix} x \\ u \end{bmatrix}.
\vspace*{-0.25\baselineskip}
\end{equation}
Further, we assume that the control inputs $u$ satisfy $u\in\mathbb{U}$, where $\mathbb{U}\subset\mathbb{R}^{n_u}$ is a user-defined compact set. 
This is needed to derive the quadratic bound~\eqref{eq:proportional-bound-residual} and is motivated by the fact that, in practice, $u$ is typically bounded, e.g., due to physical constraints.
\par
\vspace*{-0.75\baselineskip}
We first derive the quadratic error bound~\eqref{eq:proportional-bound-residual} for the individual (Section~\ref{sec:controller-design-individual-bounds}) and ellipsoidal (Section~\ref{sec:controller-design-ellipsoidal-bounds}) identification error bounds. 
Then, we present a regional and a possibly global controller design based on the quadratic error bounds derived in Section~\ref{sec:controller-design-controller}.
\vspace*{-0.75\baselineskip}
\subsection{Individual identification error bounds}\label{sec:controller-design-individual-bounds}
\vspace*{-0.75\baselineskip}
In this section, we consider the individual identification error bounds presented in Theorem~\ref{th:ApproxAll}.
The following proposition characterizes how the results in Theorem~\ref{th:ApproxAll} can be transferred to the error bound~\eqref{eq:proportional-bound-residual}.
\vspace*{-0.75\baselineskip}
\begin{prop}\label{prop:QDelta-sample-complexity}
    Consider the bilinear system~\eqref{eq:BiLinSys} and let the identification error be bounded according to~\eqref{eq:errorBoundIndiv} 
    with probability at least $1-\delta$ with $\delta\in(0,1)$.
    Then, if $u\in\mathbb{U}$, the residual $r(x,u)$ of the uncertain bilinear system~\eqref{eq:BilinSysIdentified} satisfies the quadratic bound~\eqref{eq:proportional-bound-residual} for
    \vspace*{-0.45\baselineskip}
    \begin{equation}\label{eq:QDelta-sample-complexity}
        Q_\Delta 
        = 
        \begin{bmatrix}
            2 c_x^2 I_{n_x} & 0 \\ 0 & 2 c_u^2 I_{n_u}
        \end{bmatrix}
    \vspace*{-0.45\baselineskip}
    \end{equation}
    with probability at least $1-\delta$, where 
    \begin{subequations}\label{eq:constants-cx-cu}
        \begin{align}
            \!\!\!\!
            c_x &= \Big[\max_{u\in\mathbb{U}} |1-\sum_{i=1}^{n_u}[u]_i|\Big] \varepsilon_A + \Big[\max_{u\in\mathbb{U}}\sum_{i=1}^{n_u}|[u]_i| \varepsilon_{B_i}\Big],
            \!\\\!\!\!\!
            c_u &= \sqrt{\sum\nolimits_{i=1}^{n_u}\varepsilon_{[B_0]_i}^2}.
        \end{align} 
    \end{subequations} 
\end{prop}
\vspace*{-2\baselineskip}
\begin{pf}
    See Appendix~\ref{sec:appendix:proofControlBoundsIndividual}.
\end{pf}
\vspace*{-1.95\baselineskip}
Based on the individual identification error bounds derived in Theorem~\ref{th:ApproxAll}, Proposition~\ref{prop:QDelta-sample-complexity} yields a quadratic bound on the residual.
In particular, the bound is proportional to the state and input, allowing a robust controller design since the error bound vanishes at the equilibrium $(x,u)=(0,0)$.
As shown in Section~\ref{sec:controller-design-controller}, this allows the design of a stabilizing controller for the unknown bilinear system.
\vspace*{-0.6\baselineskip}
\subsection{Ellipsoidal identification error bounds}\label{sec:controller-design-ellipsoidal-bounds}
\vspace*{-0.75\baselineskip}
Next, we use the ellipsoidal identification error bounds presented in Theorem~\ref{th:ApproxAllData} and derive a corresponding matrix $Q_\Delta$. 
Here, we consider a block-wise decomposition of the matrices $\mathcal{E}_{B_i}=\left[\begin{smallmatrix}
    [\mathcal{E}_{B_i}]_{11} & [\mathcal{E}_{B_i}]_{12} \\ 
    [\mathcal{E}_{B_i}]_{21} & [\mathcal{E}_{B_i}]_{22}
\end{smallmatrix}\right]$ in~\eqref{eq:ellipBoundsTotalB}, where $[\mathcal{E}_{B_i}]_{11}\in\mathbb{R}^{n_x\times n_x}$, $[\mathcal{E}_{B_i}]_{12}=[\mathcal{E}_{B_i}]_{21}^\top\in\mathbb{R}^{n_x}$, $[\mathcal{E}_{B_i}]_{22}\in\mathbb{R}$.
\vspace*{-0.75\baselineskip}
\begin{prop}\label{prop:QDelta-data-ellipsoidal}
    Consider the bilinear system~\eqref{eq:BiLinSys} and let the identification error be bounded according to~\eqref{eq:ellipBoundsTotal} with probability at least $1-\delta$ with $\delta\in(0,1)$.
    Then, if $u\in\mathbb{U}$, the residual $r(x,u)$ of the uncertain bilinear system~\eqref{eq:BilinSysIdentified} satisfies the quadratic bound~\eqref{eq:proportional-bound-residual} for
    \begin{subequations}\label{eq:QDelta-data-ellipsoidal}
    \begin{align}
        Q_\Delta 
        = 
        \begin{bmatrix}
            (n_u+1) \max\limits_{u\in\mathbb{U}}|1 - \sum_{i=1}^{n_u}[u]_i|^2 \mathcal{E}_A & 0 \\ 0 & 0
        \end{bmatrix}
        + (n_u+1)\hat{\mathcal{E}}_B
    \nonumber\\[-0.85\baselineskip]\\[-1.1\baselineskip]\nonumber
    \end{align}
    with probability at least $1-\delta$, where 
    \begin{align}\label{eq:tildeE_B-ellipsoidal}
        \hat{\mathcal{E}}_B
        &=
        \left[\star\right]^\top
        \tilde{\mathcal{E}}_B
        \begin{bmatrix}
            (\max\limits_{u\in\mathbb{U}} |u| \otimes I_{n_x}) & 0 \\ 
            0 & I_{n_u}
        \end{bmatrix},
        \\\nonumber
        \tilde{\mathcal{E}}_B
        &=
        \scalebox{0.75}{$
            \left[\begin{array}{ccc|ccc}
                [\mathcal{E}_{B_1}]_{11} & & & [\mathcal{E}_{B_1}]_{12} & & \\
                & \ddots & & & \ddots \\
                & & [\mathcal{E}_{B_{n_u}}]_{11} & & & [\mathcal{E}_{B_{n_u}}]_{12} \\\hline
                [\mathcal{E}_{B_1}]_{21} & & & [\mathcal{E}_{B_1}]_{22} & & \\
                & \ddots & & & \ddots \\ 
                & & [\mathcal{E}_{B_{n_u}}]_{21} & & & [\mathcal{E}_{B_{n_u}}]_{22}
            \end{array}\right]
        $}.
    \end{align}
    \end{subequations}
\end{prop}
\vspace*{-1.75\baselineskip}
\begin{pf}
    See Appendix~\ref{sec:appendix:proofControlBoundsEllipsoidal}.
\end{pf}
\vspace*{-1.6\baselineskip}
Similar to the discussion for the individual identification error bounds, Proposition~\ref{prop:QDelta-data-ellipsoidal} establishes an error characterization of the residual which is tailored to control and vanishes at the origin.
\vspace*{-0.65\baselineskip}
\subsection{Controller design}\label{sec:controller-design-controller}
\vspace*{-0.65\baselineskip}
In the following, we present the proposed controller designs for system~\eqref{eq:BiLinSys} based on the identified bilinear system~\eqref{eq:BilinSysIdentified}.
To this end, we propose two different design methods with stability guarantees in Sections~\ref{sec:controller-LMI} and~\ref{sec:controller-SOS}, where the former is based on linear robust control techniques, while the latter relies on \ac{SOS} optimization.
Depending on the practical application, the different designs offer a trade-off between control performance, the size of the \ac{RoA}, and computational complexity.
\vspace*{-0.65\baselineskip}
\subsubsection{Controller design via linear robust control}\label{sec:controller-LMI}
\vspace*{-0.65\baselineskip}
First, we make use of the state-feedback controller design presented in~\citet{strasser:berberich:allgower:2023b,strasser:schaller:worthmann:berberich:allgower:2024b}.
Here, the controller design of the uncertain bilinear system~\eqref{eq:BilinSysIdentified} is addressed by \emph{linear} robust control techniques, where we rewrite the bilinear system as a linear fractional representation~\citep{zhou:doyle:glover:1996} within a user-defined state region $\mathcal{X}\subset\mathbb{R}^{n_x}$.
To this end, we define the set 
\begin{equation}\label{eq:state-constraint-set}
    \mathcal{X} = \left\{
        x\in\mathbb{R}^{n_x} \middle| 
        \begin{bmatrix}x \\ 1\end{bmatrix}^\top
        \begin{bmatrix}
            Q_x & S_x \\ S_x^\top & R_x
        \end{bmatrix}
        \begin{bmatrix}x \\ 1\end{bmatrix}
        \geq 0
    \right\},
\end{equation}
where $Q_x\prec 0$ and $R_x > 0$. 
Then, the proposed controller design guarantees invariance of $\mathcal{X}$ and regional stability of the closed loop for initial conditions in a subset of $\mathcal{X}$.
Possible choices are, e.g., $Q_x=-I$, $S_x=0$, and $R_x=c$ defining a norm bound on the state $\|x\|^2\leq c$. 
An algorithm to heuristically optimize the geometry of $\mathcal{X}$ based on the identified system dynamics is given in~\citet[Procedure~8]{strasser:schaller:worthmann:berberich:allgower:2025}, to which we also refer for a discussion on how the resulting control behavior is affected by the choice of $\mathcal{X}$. 
Further, we assume the existence of
$
    \left[\begin{smallmatrix}
        \tilde{Q}_x & \tilde{S}_x \\ \tilde{S}_x^\top & \tilde{R}_x
    \end{smallmatrix}\right] 
    \coloneqq \left[\begin{smallmatrix}
        Q_x & S_x \\ S_x^\top & R_x
    \end{smallmatrix}\right]^{-1}
$.
\begin{thm}\label{th:controller-design}
    Consider the bilinear system~\eqref{eq:BiLinSys} and let the identification error bound~\eqref{eq:errorBoundIndiv} or~\eqref{eq:ellipBoundsTotal} hold with probability at least $1-\delta$ with $\delta \in (0,1)$.
    Find $P=P^\top\succ 0$ of size $n_x\times n_x$, $L\in\R^{n_u\times n_x}$, $L_w\in\R^{n_u\times n_xn_u}$, $\Lambda=\Lambda^\top\succ 0$ of size $n_u\times n_u$, and $\nu>0$, $\tau>0$ such that the \acp{LMI}~\eqref{eq:LMI1-Lyapunov-decay} and 
    \begin{figure*}[t]
        \small
        \begin{equation}\label{eq:LMI1-Lyapunov-decay}
            \begin{bmatrix}
                P - \tau I_{n_x}
                & -\hat{A}_{ux}(\Lambda\otimes \tilde{S}_x) -\hat{B}_0 L_w (I_{n_u}\otimes \tilde{S}_x)
                & 0
                & \hat{A} P + \hat{B}_0 L
                & \hat{A}_{ux} (\Lambda\otimes I_{n_x}) + \hat{B}_0 L_w
                \\
                \star 
                & \Lambda\otimes \tilde{R}_x - L_w(I_{n_u}\otimes \tilde{S}_x) - (I_{n_u}\otimes \tilde{S}_x^\top)L_w^\top
                & -(I_{n_x}\otimes \tilde{S}_x^\top)\left[\begin{smallmatrix}0\\L_w\end{smallmatrix}\right]
                & L 
                & L_w
                \\
                \star 
                & \star 
                & \tau Q_\Delta^{-1}
                & \left[\begin{smallmatrix}P\\L\end{smallmatrix}\right]
                & -\left[\begin{smallmatrix}0\\L_w\end{smallmatrix}\right]
                \\
                \star
                & \star
                & \star 
                & P 
                & 0 
                \\
                \star 
                & \star
                & \star 
                & \star 
                & -\Lambda \otimes \tilde{Q}_x^{-1}
            \end{bmatrix}
            \succ 0
        \end{equation}
        \normalsize
        \medskip
        \vspace*{-0.85\baselineskip}
        \hrule
        \medskip
        \vspace*{-0.85\baselineskip}
    \end{figure*}
    \vspace*{-0.25\baselineskip}
    \begin{equation}\label{eq:LMI2-invariance}
        \begin{bmatrix}
            \nu \tilde{R}_x - 1 
            & -\nu \tilde{S}_x^\top \\
            -\nu \tilde{S}_x 
            & \nu \tilde{Q}_x + P
        \end{bmatrix}
        \preceq 0
    \end{equation}
    hold. 
    If the controller
    \begin{equation}\label{eq:control-law-explicit}
        u=\kappa_\mathrm{LMI}(x) = (I - L_w(\Lambda^{-1}\otimes x))^{-1} L P^{-1} x
    \end{equation}
    satisfies $\kappa_\mathrm{LMI}(x)\in\mathbb{U}$ for all $x \in \mathcal{X}_\mathrm{LMI}\coloneqq\{x\in\mathbb{R}^{n_x} \mid x^\top P^{-1} x \leq 1\}\subseteq\mathcal{X}$, then it ensures exponential stability of the closed-loop bilinear system for all initial conditions in $\mathcal{X}_\mathrm{LMI}$ with probability at least $1-\delta$.
\end{thm}
\vspace*{-1.65\baselineskip}
\begin{pf}
    According to Propositions~\ref{prop:QDelta-sample-complexity} and~\ref{prop:QDelta-data-ellipsoidal}, if $u\in\mathbb{U}$, the residual of the system identification satisfies~\eqref{eq:proportional-bound-residual} for some $Q_\Delta$ with probability at least $1-\delta$.
    Thus, the result follows directly from~\citet[Theorem~4]{strasser:berberich:allgower:2023b} and~\citet[Theorem~4.1]{strasser:schaller:worthmann:berberich:allgower:2024b}, generalized by exploiting $Q_\Delta$ in the residual bound~\eqref{eq:proportional-bound-residual}.
\end{pf}
\vspace*{-1.65\baselineskip}
Theorem~\ref{th:controller-design} establishes a controller design for the unknown bilinear system~\eqref{eq:BiLinSys} with end-to-end guarantees based on finite stochastic data.
In particular, the controller design ensures exponential stability of the closed-loop system for all initial conditions in the \ac{RoA} $\mathcal{X}_\mathrm{LMI}$ with high probability.
To this end, we exploit the identified system dynamics and the identification error bounds derived in Section~\ref{sec:non-asymp-bounds} to establish a controller design that is robust to the residual.
Here, we can use the individual or the ellipsoidal identification error bounds to derive the matrix $Q_\Delta$ used in the controller design, see~\eqref{eq:QDelta-sample-complexity} and~\eqref{eq:QDelta-data-ellipsoidal}, respectively.
The design scheme requires and ensures that the state $x$ remains within the set $\mathcal{X}$.
Thus, the set $\mathcal{X}$ needs to be carefully chosen when applying the controller design. 
More precisely, if the controller design is not feasible for a given set of data, feasibility might be ensured by either collecting more data to reduce the identification error or adjusting the set $\mathcal{X}$ to shrink the guaranteed \ac{RoA} $\mathcal{X}_\mathrm{LMI}\subseteq\mathcal{X}$.
Further, the designed controller needs to respect the input constraints $\mathbb{U}$ for all states in the \ac{RoA}. 
If this is not the case, either $\mathbb{U}$ can be enlarged or the \ac{RoA} can be decreased by shrinking $\mathcal{X}$.
Thus, the controller design can be iterated until successful, see Algorithm~\ref{algo:DDCBilinear}.
Note that the controller design is cast as a \ac{SDP}, which can be efficiently solved. 
We observe that the computational complexity of the \acp{LMI}~\eqref{eq:LMI1-Lyapunov-decay},~\eqref{eq:LMI2-invariance} is $\mathcal{O}((3n_x + n_u (2 + n_x))^6)$.
This can be challenging for large-scale systems and, hence, future work should investigate structure-exploiting \ac{SDP} techniques~\citep{deklerk:2010,gramlich:holicki:scherer:ebenbauer:2023}.
\vspace*{-0.625\baselineskip}
\subsubsection{Controller design via \ac{SOS} optimization}\label{sec:controller-SOS}
\vspace*{-0.625\baselineskip}
Next, we propose a controller design based on \ac{SOS} optimization techniques\footnote{
We define $\mathbb{R}[x,d]$ as the set of all polynomials $s$ in the variable $x\in\mathbb{R}^n$ with real coefficients and degree at most $d$, i.e., $s(x) = \sum_{\alpha\in\mathbb{N}_0^n, |\alpha|\leq d} s_\alpha x^\alpha$, where $s_\alpha\in\mathbb{R}$ and $|\alpha|\leq d\in\mathbb{N}_0$.
Here, the monomials are given by $x^\alpha = x_1^{\alpha_1}\cdots x_n^{\alpha_n}$ for a multi-index $\alpha\in\mathbb{N}_0^n$ with $|\alpha|= \alpha_1 + \cdots + \alpha_n$.
The set of all $p\times q$-matrices whose elements belong to $\mathbb{R}[x,d]$ is denoted by $\mathbb{R}[x,d]^{p\times q}$.
A matrix $S\in\mathbb{R}[x,2d]^{p\times p}$ is called \ac{SOS} matrix in $x$ if it can be decomposed as $S = T^\top T$ for some $T\in\mathbb{R}[x,d]^{q\times p}$, where we write $S\in\SOS[x,2d]^p$.
Finally, $S\in\mathbb{R}[x,2d]^{p\times p}$ is said to be strictly \ac{SOS} if there exists $\varepsilon>0$ such that $(S-\varepsilon I_p)\in\SOS[x,2d]^p$, denoted by $S\in\SOS_+[x,2d]^p$.}.
The main motivation for this is that choosing the pre-defined set $\mathcal{X}$ in~\eqref{eq:state-constraint-set} may be restrictive for some applications.
Instead, we aim to design a rational controller $u=\kappa_\mathrm{SOS}(x)$ that ensures closed-loop stability with high probability without relying on $\mathcal{X}$.
To this end, we adapt the \ac{SOS} optimization-based design proposed in~\citet{strasser:berberich:allgower:2025} to handle the identification error bound~\eqref{eq:proportional-bound-residual} with matrix $Q_\Delta$.
\vspace*{-0.75\baselineskip}
\begin{thm}\label{th:controller-design-SOS}
    Consider the bilinear system~\eqref{eq:BiLinSys} and let the identification error bound~\eqref{eq:errorBoundIndiv} or~\eqref{eq:ellipBoundsTotal} hold with probability at least $1-\delta$ with $\delta \in (0,1)$.
    Find $\alpha>0$, $P=P^\top\succ 0$ of size $n_x\times n_x$, $L_\mathrm{n}\in\R[x,2\alpha-1]^{n_u\times n_x}$, $\tau\in\SOS_+[x,2\alpha]$, $u_\mathrm{d}\in\SOS_+[x,2\alpha]$, and $\rho > 0$ such that
    \begin{equation}\label{eq:controller-design-SOS}
        \hspace*{-0.035\linewidth}
        \begin{bmatrix}
            u_\mathrm{d} P - \tau I_{n_x}
            & 0
            & u_\mathrm{d} \hat{A} P + \hat{B}_0 L_\mathrm{n} + \hat{A}_{ux} (L_\mathrm{n} \otimes x)
            \\
            \star
            & \tau Q_\Delta^{-1}
            & \left[\begin{smallmatrix}u_\mathrm{d} P \\ L_\mathrm{n} \end{smallmatrix}\right]
            \\
            \star
            & \star
            & u_\mathrm{d} (P-\rho I_{n_x})
        \end{bmatrix}
        \!\!\!
    \end{equation}
    is in $\SOS[x,2\alpha]^{3n_x + n_u}$.
    Then, the controller
    \begin{equation}\label{eq:control-law-SOS}
        u=\kappa_\mathrm{SOS}(x) = \tfrac{1}{u_\mathrm{d}(x)} L_\mathrm{n}(x) P^{-1} x
    \end{equation}
    ensures exponential stability of the closed-loop bilinear system for all initial conditions in $\mathcal{X}_\mathrm{SOS}(c^*)$ with probability at least $1-\delta$, where $\mathcal{X}_\mathrm{SOS}(c)\coloneqq\{x\in\mathbb{R}^{n_x} \mid x^\top P^{-1} x \leq c\}$ and $c^*=\argmax \{c \in \mathbb{R}_+\cup \{\infty\}\mid \kappa_\mathrm{SOS}(x)\in\mathbb{U} \text{ for all } x\in\mathcal{X}_\mathrm{SOS}(c)\}$.
\end{thm}
\vspace*{-1.85\baselineskip}
\begin{pf}
    This is a direct consequence of~\citet[Theorem~4]{strasser:berberich:schaller:worthmann:allgower:2025}, generalized by exploiting $Q_\Delta$ in the residual error bound~\eqref{eq:proportional-bound-residual}.
\end{pf}
\vspace*{-1.85\baselineskip}
\begin{algorithm}[t]
    \caption{Indirect DDC with end-to-end guarantees}
    \bgroup
    \begin{algorithmic}[1]
        \State Choose $\delta$, $T_0,\dots, T_{n_u}$, $\mathbb{U}$
        \State Collect data and identify system matrices using Algorithm~\ref{algo:IDSampling} with the sampling scheme chosen according to the desired error bounds
        \State Attempt controller design:
        \Statex \quad\emph{a) LMI-based controller design (Theorem~\ref{th:controller-design})}
        \makeatletter
        \renewcommand{\alglinenumber}[1]{\arabic{ALG@line}a:}
        \makeatother
        \State Choose $\mathcal{X}$ in~\eqref{eq:state-constraint-set}
        \If{\acp{LMI}~\eqref{eq:LMI1-Lyapunov-decay} and~\eqref{eq:LMI2-invariance} are feasible}
            \State Controller~\eqref{eq:control-law-explicit} yields closed-loop exponential
            \Statex \quad\;\;stability in $\mathcal{X}_\mathrm{LMI}\subseteq\mathcal{X}$ with high probability
        \Else
            \State Modify parameters in 1 until successful
        \EndIf
        
        \Statex \quad\emph{b) SOS-based controller design (Theorem~\ref{th:controller-design-SOS})}
        \addtocounter{ALG@line}{-6}
        \makeatletter
        \renewcommand{\alglinenumber}[1]{\arabic{ALG@line}b:}
        \makeatother
        \If{SOS condition~\eqref{eq:controller-design-SOS} is feasible}
            \State Controller~\eqref{eq:control-law-SOS} yields closed-loop exponential
            \Statex \quad\;\;stability in $\mathcal{X}_\mathrm{SOS}(c^*)$ with high probability
        \Else
            \State Modify parameters in 1 until successful
        \EndIf
    \end{algorithmic}
    \egroup
    \label{algo:DDCBilinear}
\end{algorithm}

Theorem~\ref{th:controller-design-SOS} establishes a controller design for the unknown bilinear system~\eqref{eq:BiLinSys} with end-to-end guarantees based on finite stochastic
data for all states in the \ac{RoA} $\mathcal{X}_\mathrm{SOS}$.
If the quadratic bound~\eqref{eq:proportional-bound-residual} would hold globally,~\citet[Theorem~2]{strasser:berberich:allgower:2025} directly yields \emph{global} stability guarantees for the unknown bilinear closed-loop system.
Although the identification error bounds~\eqref{eq:errorBoundIndiv} and~\eqref{eq:ellipBoundsTotal} do hold globally, the subsequently established residual error bounds in Propositions~\ref{prop:QDelta-sample-complexity} and~\ref{prop:QDelta-data-ellipsoidal} hold only for $u\in\mathbb{U}$ and, thus, only in a certain region of the state space around the equilibrium.
While future research should investigate how to ensure global residual error bounds, we restrict the designed controller here to $u\in\mathbb{U}$ and we provide regional stability guarantees for initial conditions in $\mathcal{X}_\mathrm{SOS}$.
The established \ac{RoA} $\mathcal{X}_\mathrm{SOS}$ is typically significantly larger than $\mathcal{X}_\mathrm{LMI}$ and may even be global (i.e., $c^*=\infty$) for certain systems and input sets $\mathbb{U}$, see the numerical examples in Section~\ref{sec:numerics-controller}.
The reason for this is that $\kappa_\mathrm{SOS}$ is a rational controller, where both the numerator and denominator share the same degree.
Thus, the controller does not grow unboundedly for small or large states.
Further, we note that the denominator of $\kappa_\mathrm{SOS}$ has no zeros due to the strict SOS property of $u_\mathrm{d}$.
Overall, the two controller designs offer a trade-off between control performance, the size of the \ac{RoA}, and computational complexity.
Note that the \ac{SOS} program~\eqref{eq:controller-design-SOS} is linear in the decision variables $P$, $L_\mathrm{n}$, $\tau$, $\rho$ for a fixed polynomial $u_\mathrm{d}$, and can be solved using convex optimization techniques~\citep{papachristodoulou:prajna:2005}.
Here, the a priori chosen $u_\mathrm{d}$ acts as a tuning parameter, providing an additional degree of freedom in the controller parametrization.
Compared to Theorem~\ref{th:controller-design}, the \ac{SOS}-based design in Theorem~\ref{th:controller-design-SOS} does not rely on a pre-defined state region $\mathcal{X}$ and, hence, is applicable to a wider range of systems.
On the other hand, the \ac{SOS} optimization with complexity $\mathcal{O}((3n_x+n_u)^6 n_x^{6\alpha})$ is computationally more demanding and, hence, might not be suitable for large-scale systems.
Moreover, a higher degree $\alpha$ in the controller design allows for a more flexible controller design, but it also increases the computational complexity of the \ac{SOS} program. 
\par
\vspace*{-0.75\baselineskip}
For the controller design, we assume that the system is not affected by noise.
However, when the \ac{SOS}-based controller \emph{globally} exponentially stabilizes the nominal bilinear system, this, by~\citet[Corollary 2]{culbertson2023input}, implies exponential \ac{ISS} in probability. 
Thus, the trajectories of the perturbed closed loop remain bounded with high probability when the process noise is present during operation~\citep[Corollary 3]{culbertson2023input}.
\citet{culbertson2023input} also note that the choice of the Lyapunov function is crucial to reduce conservatism of the \ac{ISS} guarantees.
To this end, it may be beneficial to incorporate the  noise directly in the controller design via an additional uncertainty channel, and we defer this controller design improvement to future research.
%
%
\vspace*{-0.85\baselineskip}
\section{Numerical examples}\label{sec:numerics}
\vspace*{-0.85\baselineskip}
In this section, we first provide numerical simulations to illustrate the derived identification error bounds of Section~\ref{sec:non-asymp-bounds} (Section~\ref{sec:numerics-error-bounds}), where we compare the a priori identification error bounds (Theorem~\ref{th:sampleCompelxitySubG}) with the data-dependent bounds (Theorem~\ref{co:ThwithMFromData}). 
Second, we use both types of identification error bounds to design a controller for a bilinear system providing end-to-end guarantees for indirect data-driven control (Section~\ref{sec:numerics-controller}).\footnote{The code for the numerical examples can be accessed at:~\href{https://github.com/col-tasas/2024-bilinear-end-to-end}{https://github.com/col-tasas/2024-bilinear-end-to-end}}
\vspace*{-0.75\baselineskip}
\subsection{Error bounds}\label{sec:numerics-error-bounds}
\vspace*{-0.75\baselineskip}
In this section, we analyze the identification error bounds derived in Section~\ref{sec:non-asymp-bounds} with respect to conservatism and dependence on key problem parameters.
Here, we focus our analysis on the identification problem~\eqref{eq:BiLinSys2} and refer to the works by~\citet{dean2020sample} or~\citet{matni2019tutorial} for the analysis of the identification error bounds of~\eqref{eq:BiLinSys1}.
For the remainder of this section the setup will be as follows.
Data is collected by sampling $x^{(\ell)}\simiid\N (0, I)$ and evaluating the unknown, affine function 
\begin{equation*}
    x_+ = B_1 x + B_0 + w,
\vspace*{-0.25\baselineskip}
\end{equation*}
where $w_t\simiid \N(0, \sigma_w^2 I_{n_x})$, $\sigma_w = 0.5$. 
The matrix $B_1\in \R^{n_x\times n_x}$ and the vector $B_0\in\R^{n_x}$ are drawn randomly. 
We estimate $B_1$ and $B_0$ using the \ac{OLS} estimate~\eqref{eq:LsSol2}. 
\par
\vspace*{-0.75\baselineskip}
First, we consider the influence of the sample size on the identification error. 
To this end, we select $n_x=25$. 
Further, we empirically estimate the identification error through Monte Carlo simulations to average out the randomness in the noise and data-sampling.
Since the data-dependent identification error bound~\eqref{eq:dataErrorBi} also depends on the observed data, we consider the mean over the Monte Carlo simulations.
The mean and $3 \sigma$-band of the empirically approximated identification error as well as the mean of the data-based bound are displayed in Fig.~\ref{fig:Comp-Bounds-T-a} along with the a priori sample complexity bounds.
\begin{figure*}[t]
    \captionsetup[subfloat]{captionskip=0ex}
    \centering
    \subfloat[Error for different sample sizes.]{\label{fig:Comp-Bounds-T-a}
        \resizebox{0.4655\linewidth}{!}{\input{figures/compBounds/errorCompSamples_nx25_T25000_delta005_nMC100.tex}}
    }
    \hfill
    \subfloat[Error times $\sqrt{T}$ over sample size.]{\label{fig:Comp-Bounds-T-b}
        \resizebox{0.4655\linewidth}{!}{\input{figures/compBounds/errorCompSamples_nx25_T25000_delta005_nMC100_rel.tex}}
    }
    \\[-0.75\baselineskip]
    \subfloat[Error for different problem dimensions.]{\label{fig:Comp-Bounds-nx-a}
        \resizebox{0.4655\linewidth}{!}{
%
%

\begin{tikzpicture}
\definecolor{oiBlue}{HTML}{0072B2}
\definecolor{oiOrange}{HTML}{E69F00}

\begin{axis}[%
width=7.845cm,
height=2cm,
at={(0in,0in)},
scale only axis,
xmin=1,
xmax=30,
xlabel style={font=\color{white!15!black}},
xlabel={$n_x$},
ymin=0,
ymax=0.125,
ylabel style={font=\color{white!15!black}},
ylabel={$\Vert \hat B_1- B_1\Vert_2$},
axis background/.style={fill=white},
xtick ={1, 5, 10, 15, 20, 25, 30},
xticklabels= {$1$, $5$, $10$, $15$, $20$, $25$, $30$},
ytick={0, 0.05, 0.1},
minor y tick num=1,
yticklabels={$0$, $0.025$, $0.05$, $0.075$, $0.1$, $0.125$},
grid=both,
]
\addplot [color=black, line width=1pt, mark=asterisk, mark options={solid, mark size=3pt,black}]
  table[row sep=crcr]{%
1	0.029535547127775\\
2	0.0347359916986512\\
3	0.0392779681632262\\
4	0.0433752968559029\\
5	0.0471416723357316\\
7	0.0539752352597623\\
10	0.0629608269174844\\
15	0.0758115131583754\\
20	0.0870404531460364\\
25	0.0971187189537454\\
30	0.106389065787709\\
};
\label{fig:Ex2_Priori}

\addplot[area legend, draw=none, fill=white!80!oiOrange, forget plot]
table[row sep=crcr] {%
x	y\\
1	0.0135658946074828\\
2	0.016223679426291\\
3	0.0184912050264622\\
4	0.0205213281915187\\
5	0.0223752671528524\\
7	0.0257188907002616\\
10	0.0300435142562769\\
15	0.0361723524488227\\
20	0.0414413691241432\\
25	0.0461395352228758\\
30	0.0504207600372032\\
30	0.0508563232436365\\
25	0.046530244478056\\
20	0.0418287874522928\\
15	0.0365086445168134\\
10	0.0303524429115382\\
7	0.0259874051147358\\
5	0.0226547561738558\\
4	0.0207653319833944\\
3	0.0187081179905433\\
2	0.0164108237610935\\
1	0.0137245383036232\\
}--cycle;
\addplot [color=oiOrange, line width=1pt, mark=asterisk, mark options={solid, mark size=3pt, oiOrange}]
  table[row sep=crcr]{%
1	0.0136452164555543\\
2	0.0163172515936907\\
3	0.0185996615085031\\
4	0.0206433300874558\\
5	0.022515011663355\\
7	0.0258531479074975\\
10	0.0301979785839066\\
15	0.0363404984828186\\
20	0.0416350782882162\\
25	0.046334889850467\\
30	0.0506385416404199\\
};
\label{fig:Ex2_Data}

\addplot[area legend, draw=none, fill=white!80!oiBlue, forget plot]
table[row sep=crcr] {%
x	y\\
1	-0.00102246366124154\\
2	-0.000168392289147355\\
3	0.000789018409750249\\
4	0.00102446531721353\\
5	0.00176089028662127\\
7	0.00270328023477546\\
10	0.00387854258642675\\
15	0.00567867202583091\\
20	0.00685048485874037\\
25	0.00799034504181007\\
30	0.00910457244608032\\
30	0.0118725588551663\\
25	0.0110263028223996\\
20	0.00983942989629847\\
15	0.00861911142678186\\
10	0.00744475549815558\\
7	0.00646432255762144\\
5	0.0056356293164839\\
4	0.00517637020828774\\
3	0.00441160154020118\\
2	0.00355279686972978\\
1	0.00254405402959312\\
}--cycle;
\addplot [color=oiBlue, line width=1pt, mark=asterisk, mark options={solid,  mark size=3pt,oiBlue}]
  table[row sep=crcr]{%
1	0.000760795184174157\\
2	0.00169220229028966\\
3	0.00260030997497651\\
4	0.00310041776275227\\
5	0.00369825980155269\\
7	0.00458380139619763\\
10	0.00566164904229183\\
15	0.00714889172630606\\
20	0.0083449573775205\\
25	0.00950832393210632\\
30	0.0104885656506233\\
};
\label{fig:Ex2_Monte}

\end{axis}
\end{tikzpicture}%
}
    }
    \hfill
    \subfloat[Error bounds relative to mean Monte Carlo error.]{\label{fig:Comp-Bounds-nx-b}
        \resizebox{0.4655\linewidth}{!}{
%
%

\begin{tikzpicture} 
    \definecolor{oiBlue}{HTML}{0072B2}
    \definecolor{oiOrange}{HTML}{E69F00}   

    \begin{axis}[%
    width=7.845cm,
    height=2cm,
    at={(0in,0in)},
    scale only axis,
    xmin=1,
    xmax=30,
    xlabel style={font=\color{white!15!black}},
    xlabel={$n_x$},
    xtick ={1, 5, 10, 15, 20, 25, 30},
    xticklabels= {$1$, $5$, $10$, $15$, $20$, $25$, $30$},
    ymin=2.5,
    ymax=30,
    ytick = {10, 20, 30},
    minor ytick = {5, 15, 25},
    ylabel style={font=\color{white!15!black}},
    ylabel={$\frac{\varepsilon_{B_1}}{\Vert \hat B_1- B_1\Vert_2}$},
    axis background/.style={fill=white},
    grid=both,
    ]
    \addplot [color=black, line width=1pt, mark=asterisk, mark options={solid, mark size=3pt, black}]
      table[row sep=crcr]{%
    1	38.8219428068179\\
    2	20.527091765532\\
    3	15.10510998351\\
    4	13.9901459013127\\
    5	12.746987736216\\
    7	11.7752124480189\\
    10	11.1205810263374\\
    15	10.6046525896322\\
    20	10.4303052979654\\
    25	10.2140734420947\\
    30	10.1433379292798\\
    };
    
    \addplot [color=oiOrange, line width=1pt, mark=asterisk, mark options={solid, mark size=3pt, oiOrange}]
      table[row sep=crcr]{%
    1	17.9354663901238\\
    2	9.64261287631529\\
    3	7.15286319227247\\
    4	6.65824145877108\\
    5	6.08800162008682\\
    7	5.64011083223192\\
    10	5.33377790787382\\
    15	5.08337514038615\\
    20	4.98924996314292\\
    25	4.87308701105736\\
    30	4.82797584790924\\
    };

    \end{axis}
    
    \end{tikzpicture}%
}
    }
    \vspace*{-0.5\baselineskip}
    \caption{Mean of the identification error~\eqref{fig:Ex1_Monte}, data-based bounds~\eqref{fig:Ex1_data}, and a priori error bounds~\eqref{fig:Ex1_priori} through $100$ Monte Carlo simulations. Shaded areas are respective $3\sigma$-bands.}
    \label{fig:Comp-Bounds}
    \vspace*{-0.5\baselineskip}
\end{figure*}
The results show that the a priori identification error bounds are more conservative than their data-dependent counterparts. 
This is to be expected since the data-dependent identification error bounds only take into account the data that is observed and need less potentially conservative steps in their derivation. 
Further, even in a high-data regime, both bounds overestimate the true identification error, where as expected the absolute value of the gap decreases as the number of samples increases. 
Additionally, Fig.~\ref{fig:Comp-Bounds-T-b} shows that the dependency $\bigO(\nicefrac{1}{\sqrt{T}})$ on the sample size captured in the a priori identification error bound is correct.
\par  
\vspace*{-0.75\baselineskip}
Now, we focus on the dependence on the problem dimension $n_x$. 
To this end, we consider random $B_1$ and $B_0$ with $n_x$ varying between $1$ and $30$.
Again, we use Monte Carlo simulations to evaluate the identification errors and identification error bounds for $T=\SI{250000}{}$ samples. The results are displayed in Fig.~\ref{fig:Comp-Bounds-nx-a}.
It is apparent that the identification error increases as expected with the problem size. 
However, as shown in Fig.~\ref{fig:Comp-Bounds-nx-b}, the conservatism of the error bounds decreases as the problem size $n_x$ increases.
As before, the data-based bound is consistently less conservative than the a priori bound.  
%
%
\vspace*{-0.75\baselineskip}
\subsection{Controller design}\label{sec:numerics-controller}
\vspace*{-0.75\baselineskip}
Next, we study the incorporation of the different types of identification error bounds for control design. 
In particular, we design a controller for the bilinear system~\eqref{eq:BiLinSys} following Theorems~\ref{th:controller-design} and~\ref{th:controller-design-SOS} with the data-dependent individual identification error bounds (Section~\ref{sec:controller-design-individual-bounds}) and the ellipsoidal identification error bounds (Section~\ref{sec:controller-design-ellipsoidal-bounds}).
Here, we study the different error bounds in terms of 1) the feasibility of the controller designs depending on the data length, and 2) the size of the guaranteed \ac{RoA} of the regional design in Theorem~\ref{th:controller-design}.
We note that we do not consider the a priori identification error bounds in the controller design, as they are more conservative than the data-dependent bounds according to the previous section.
The simulations are performed in MATLAB using the YALMIP toolbox~\citep{lofberg:2004} with its \ac{SOS} module~\citep{lofberg:2009} and the solver~\citet{mosek:2022}.
\vspace*{-0.75\baselineskip}
\subsubsection{Academic example}\label{sec:exmp-2D-academic}
\vspace*{-0.75\baselineskip}
We start by considering the dynamical system 
\begin{equation}
    x_+ = \begin{bmatrix}
        1 & 1 \\ 
        0 & 1
    \end{bmatrix}
    x 
    + 
    \begin{bmatrix}
        1 \\ 1
    \end{bmatrix}
    u 
    + 
    \begin{bmatrix}
        1 & 0 \\
        0 & 1
    \end{bmatrix}
    u x
    + w,
\end{equation} 
where we assume a compact input space $\mathbb{U}=[-2,2]$.
For the \emph{regional} controller design in Section~\ref{sec:controller-LMI}, we define the region of interest $\mathcal{X}$ in~\eqref{eq:state-constraint-set} via the norm bound $\|x\|_2^2\leq c$.
We emphasize that with high probability the \ac{RoA} $\mathcal{X}_\mathrm{LMI}\subseteq\mathcal{X}$ is invariant under our control law~\eqref{eq:control-law-explicit}, where $\mathcal{X}_\mathrm{LMI}$ corresponds to the sublevel set $V(x)\leq 1$ of the Lyapunov function $V(x) = x^\top P^{-1} x$ for $P\succ 0$.
Hence, we maximize the trace of $P$ subject to the \acp{LMI}~\eqref{eq:LMI1-Lyapunov-decay},~\eqref{eq:LMI2-invariance} to find the largest \ac{RoA}.
In comparison, the SOS design in Section~\ref{sec:controller-SOS} is performed for $\alpha=1$ and $u_\mathrm{d}(x)=1+[x]_1^2 + [x]_1[x]_2 + [x]_2^2\in\SOS_+[x,2\alpha]$.
\par  
\vspace*{-0.75\baselineskip}
First, we study the feasibility of the controller designs for the two types of identification error bounds. 
In particular, we select $w \simiid \mathcal{N}(0, \sigma_w^2I)$ with $\sigma_w=\SI{0.1}{}$, $\delta=0.05$, sample $x^{(\ell)}\simiid \mathcal{N}(0, I)$, and numerically determine the minimally required data lengths. 
Here, we restrict ourselves to data lengths $T_0=T_1=T$, but note that generally the lengths could be chosen differently. 
\begin{table}[t]
    \vspace*{-0.5\baselineskip}
    \caption{Required data length for a feasible controller design for the academic example in Section~\ref{sec:exmp-2D-academic}.}
    \label{tab:exmp-2D-feasibility-2D-academic}
    \centering
    \resizebox{0.9\columnwidth}{!}{
        \begin{tabular}{c|ccc|c}
            \cmidrule[\heavyrulewidth]{2-5}
            & \multicolumn{3}{c|}{$\kappa_\mathrm{LMI}(x)$, \ac{RoA}: $\|x\|_2^2 \leq c$} & $\kappa_\mathrm{SOS}(x)$
            \\
            & $c=0.1$ & $c=0.6$ & $c=0.9$ & RoA: $\mathbb{R}^2$ \\
            \midrule
            Indiv. & $T=\SI{360}{}$ & $T=\SI{2263}{}$ & $T=\SI{34668}{}$ & $T=\SI{5145}{}$ \\
            Ellips. & $T=\SI{33}{}$ & $T=\SI{213}{}$ & $T=\SI{3999}{}$ & $T=\SI{1040}{}$ \\
            \midrule
            \makecell{Comp.\\time} & $\SI{0.0055}{\second}$ & $\SI{0.0066}{\second}$ & $\SI{0.0070}{\second}$ & $\SI{0.0135}{\second}$ \\
            \bottomrule
        \end{tabular}
    }
    \vspace*{-0.25\baselineskip}
\end{table}
The results in Table~\ref{tab:exmp-2D-feasibility-2D-academic} show that the ellipsoidal error bounds require less data to design a feasible controller compared to the individual error bounds.
In other words, the ellipsoidal error bounds allow for larger \ac{RoA} for a given data length.
This is in line with Fig.~\ref{fig:exmp-2D-RoA}, where we show the \acp{RoA} corresponding to the simulations in Table~\ref{tab:exmp-2D-feasibility-2D-academic}.
We emphasize that the \ac{RoA} of the regional controller design is determined by $P$ whose trace is maximized in the controller design and, thus, the obtained \ac{RoA} for more data is not necessarily a superset of the \ac{RoA} for less data. 
However, this could be ensured by adding a suitable set-containment constraint to the optimization problem, which would come at the cost of less degrees of freedom in the controller design.
\begin{figure}[tb]
    \centering
    \vspace*{-0.85\baselineskip}
    \input{figures/controller/exmp-2D-RoA.tex}
    \vspace*{-\baselineskip}
    \caption{\ac{RoA} of the academic example for $\mathcal{X}=\{x\mid\|x\|^2\leq c\}$ and the minimum required data length for individual error bounds~\eqref{fig:exmp-2D-RoA-individual-minimal} and ellipsoidal error bounds~\eqref{fig:exmp-2D-RoA-ellipsoidal-minimal} as well as the \ac{RoA} with ellipsoidal error bounds for the minimum data length required for individual error bounds~\eqref{fig:exmp-2D-RoA-ellipsoidal-T-individual}.}
    \label{fig:exmp-2D-RoA}
    \vspace*{-0.25\baselineskip}
\end{figure}
Notably, the \ac{SOS} controller $\kappa_\mathrm{SOS}$ is globally stabilizing, i.e., on the entire state space $\mathcal{X}_\mathrm{SOS}(c^*) = \mathbb{R}^2$, and it is feasible for less data than the regional controller for $\|x\|_2^2\leq 0.9$.
However, the complexity of the \ac{SOS} controller design is higher than the regional controller design, which is reflected in the computation times in Table~\ref{tab:exmp-2D-feasibility-2D-academic}.
\vspace*{-0.75\baselineskip}
\subsubsection{Nonlinear inverted pendulum}\label{sec:exmp-nonlinear-pendulum}
\vspace*{-0.75\baselineskip}
Next, we illustrate the application of our results to a nonlinear system using the link with Koopman operator theory discussed in Section~\ref{sec:koop}.
To this end, we consider the nonlinear inverted pendulum~\citep{tiwari:nehma:lusch:2023,verhoek:abbas:toth:2023,strasser:schaller:worthmann:berberich:allgower:2025}
\begin{equation}
    z_+
    = \begin{bmatrix}
        [z]_1 + T_s [z]_2 \\
        [z]_2 + \frac{T_sg}{l}\sin([z]_1) - \frac{T_sb}{ml^2}[z]_2 + \frac{T_s}{ml^2} u
    \end{bmatrix}
\end{equation}
with mass $m=\SI{1}{\kilogram}$, length $l=\SI{1}{\meter}$, damping coefficient $b=\SI{0.5}{}$, gravitational acceleration $g=\SI{9.81}{\meter\per\second^2}$ and $T_s=\SI{0.1}{\second}$.
We follow Section~\ref{sec:koop} and define the lifting function $x=\Phi(z) = \begin{bmatrix}
    [z]_1 & [z]_2 & \sin([z]_1)
\end{bmatrix}^\top$ leading to an uncertain bilinear system representation in $x$.
Here, we assume that the lifting function $\Phi$ gives an exact finite-dimensional Koopman representation of the system dynamics.
However, we emphasize that, when an exact lifting is unknown, the approximation error of the (finite-dimensional) Koopman representation cannot be neglected in the controller design; see~\citet{strasser:berberich:schaller:worthmann:allgower:2025,strasser:schaller:berberich:worthmann:allgower:2025} for a detailed discussion.
For the simulation, we select $w \simiid \mathcal{N}(0, \sigma_w^2I)$ with $\sigma_w=\SI{0.001}{}$, $\delta=0.05$, sample $z_{t}\simiid \mathcal{N}(0, I)$.
Then, we assume a compact input space $\mathbb{U}=[-80,80]$ and collect $T_0=T_1=\SI{2000}{}$ data samples.
Following Theorem~\ref{th:ApproxAllData} and Algorithm~\ref{algo:DDCBilinear}, we determine the corresponding ellipsoidal error bounds and the respective $Q_\Delta$ in~\eqref{eq:QDelta-data-ellipsoidal}. 
For the \ac{SOS}-based controller design in Theorem~\ref{th:controller-design-SOS}, we choose $\alpha=1$ and
\begin{equation*}
    u_\mathrm{d}(x)=1+\sum_{i=0}^{2\alpha}\sum_{j=0}^{2\alpha-i} [x]_1^i [x]_2^j [x]_3^{2\alpha - i - j}
    \in\SOS_+[x,2\alpha].
\end{equation*}
The obtained closed-loop trajectories are depicted in Fig.~\ref{fig:exmp-pendulum-SOS} to illustrate the controller's performance. 
\begin{figure}[tb]
    \centering
    \input{figures/controller/exmp-pendulum-SOS.tex}
    \vspace*{-\baselineskip}
    \caption{Closed-loop trajectories of the nonlinear inverted pendulum example in Section~\ref{sec:exmp-nonlinear-pendulum} with the \ac{SOS} controller $\kappa_\mathrm{SOS}$.}
    \label{fig:exmp-pendulum-SOS}
    \vspace*{-0.25\baselineskip}
\end{figure}
Overall, these results show that our approach can be readily applied to nonlinear systems using the link with Koopman operator theory, paving the way towards the design of data-driven controllers for nonlinear systems using data affected by stochastic noise.
%
%
\vspace*{-0.95\baselineskip}
\section{Conclusion}\label{sec:conclusion}
\vspace*{-0.95\baselineskip}
In this work, we leveraged tools from statistical learning theory to derive finite sample identification error bounds for the identification of unknown bilinear systems from noisy data. 
The derived identification error bounds are then combined with robust control for bilinear systems to obtain an exponentially stable closed loop. 
The presented numerical studies show the interplay between identification error bounds and controller design.
To the best of our knowledge, this is the first work connecting statistical learning theory results with robust control to provide end-to-end guarantees for indirect data-driven control of bilinear systems from finite data affected by potentially unbounded stochastic noise. 
We note that the results of this work provide a promising avenue for indirect data-driven control of nonlinear systems by means of Koopman operator theory and view this as an interesting direction for future work. 

\vspace*{-0.75\baselineskip}
\bibliography{./references.bib}

\vspace*{-0.5\baselineskip}
\appendix
%
%
\vspace*{-0.5\baselineskip}
\section{Technical results for the proof of Theorem~\ref{th:sampleCompelxitySubG}} \label{app:technical_results}
\vspace*{-0.75\baselineskip}
In the following, we provide technical results used throughout the proof of Theorems~\ref{th:ID_A} and~\ref{th:sampleCompelxitySubG}.
The following proposition is a generalization of~\citet[Proposition~III.1]{matni2019tutorial} to sub-Gaussian noise and sampling. 
\vspace*{-\baselineskip}
\begin{prop} \label{prop:noiseTermSubG}
    Let $x^{(\ell)}\stackrel{\text{i.i.d}}{\sim} \subGvec{\sigma_x^2}{n_x}$ and $w^{(\ell)} \stackrel{\text{i.i.d}}{\sim} \subGvec{\sigma_w^2}{n_x}$. Fix a failure probability $\delta \in (0,1)$ and let $T_i \ge \frac{1}{2} \log\left(9^{2n_x}/\delta\right)$, then it holds that
    \begin{equation*}
        \Prob\left[\Big\Vert \sum_{\ell=1}^{T_i} x^{(\ell)} {w^{(\ell)}}^\top \Big\Vert_2 \le 4 \sigma_x \sigma_w \sqrt{T_i \log\left(9^{2n_x}/\delta\right)}\right] \le 1- \delta.
    \end{equation*}
\end{prop}
\vspace*{-\baselineskip}
\vspace*{-1.25\baselineskip}
\begin{pf}
    The proof can be carried out in the same way as the proof of \citet[Proposition~III.1]{matni2019tutorial}.
\end{pf}
\vspace*{-1.5\baselineskip}
The following Proposition provides a lower bound on the smallest eigenvalue of a covariance-like matrix and is a variant of \citet[Theorem 5.39]{vershynin2010introduction}.
\vspace*{-0.75\baselineskip}
\begin{prop} \label{prop:lowerboundCov}
   Let $x^{(\ell)}\stackrel{\text{i.i.d}}{\sim} \subGvec{\sigma_x^2}{n_x}$. 
    Fix a failure probability $\delta\in (0, 1)$ and some $c\in(0,\frac12)$. 
    Let $T_i \ge \frac{8}{c^2} \log(2\cdot 9^{n_x}/\delta) $. 
    Then it holds that 
    \begin{equation*}
        \Prob\left[\lambda_\mathrm{min}\left(\sum_{\ell=1}^{T_i} x^{(\ell)} {x^{(\ell)}}^\top \right) \ge \sigma_x T_i (1- 2c)^2\right] \ge 1- \delta .
    \end{equation*}
\end{prop}
\vspace*{-\baselineskip}
\vspace*{-1.25\baselineskip}
\begin{pf}
    First we define $\sigma_x y^{(\ell)} = x^{(\ell)}$, such that $y^{(\ell)} \stackrel{\text{i.i.d.}}{\sim} \subGvec{1}{n_x}$. 
    Now note that 
    $
        \sum_{t=1}^{T_i} x^{(\ell)} {x^{(\ell)}}^\top = \sigma_x^2 \sum_{t=1}^{T_i} y^{(\ell)} {y^{(\ell)}}^\top
    $,
    i.e., it suffices to analyze the smallest eigenvalue of $ Z = \sum_{t=1}^{T_i} y^{(\ell)} {y^{(\ell)}}^\top$.
    Next, observe that 
    $
        Z = Y^\top Y
    $,
    with $Y = \begin{bmatrix}
        y^{(1)} &
        \cdots   &
        y^{(T_i)}
    \end{bmatrix}^\top$. Using this definition it holds that $\sigma_\mathrm{min}^2(Y) = \lambda_\mathrm{min}(Z)$, where $\sigma_\mathrm{min}(Y)$ denotes the smallest singular value of $Y$.
    Applying \citet[Lemma 5.36]{vershynin2010introduction} with $B = \frac{1}{\sqrt{T_i}}Y$ yields
    \begin{equation}\label{eq:implication}
        \left\Vert \tfrac{1}{T_i} Y^\top Y - I \right\Vert \le \max(\varepsilon, \varepsilon^2) \implies \lambda_\mathrm{min}(Z) \ge T (1 - \epsilon)^2 .
    \end{equation} 
    In the following, we deploy an $\epsilon$-net argument to show the bound on the l.h.s. of~\eqref{eq:implication}. 
    To this end, let $\{v_k\}_{k=1}^M$ be an $\frac14$-covering of $\sph^{n_x-1}$. 
    By~\citet[Lemma 5.2]{vershynin2010introduction} we have $M\leq 9^{n_x}$ and consequently 
    \begin{align*}
        \left\Vert \tfrac{1}{T_i} Y^\top Y - I \right\Vert 
        &\leq 2\max_{k \in \mathbb{N}_{[1, 9^{n_x}]}} \Big\vert v_k^\top \left(\tfrac{1}{T_i} Y^\top Y - I\right) v_k \Big\vert 
        \\ 
        &= 2\max_{k \in \mathbb{N}_{[1, 9^{n_x}]}} \Big\vert \tfrac{1}{T_i} v_k^\top Y^\top Y v_k - 1 \Big\vert . 
    \end{align*} 
    Clearly, for any integer $k \in \mathbb{N}_{[1, 9^{n_x}]}$ we have $v_k^\top Y^\top Y  v_k= \sum_{\ell=1}^{T_i} v_k ^\top y^{(\ell)} {y^{(\ell)}}^\top v_k = \sum_{\ell=1}^{T_i} \zeta_\ell^2$, where $\zeta_\ell = v_k ^\top y^{(\ell)}$ follows a sub-Gaussian distribution with variance proxy $\sigma_\zeta^2 = 1$. 
    Hence, $\zeta_\ell^2$ is sub-exponentially distributed with parameters $(4, 4)$, the sum $\sum_{\ell=1}^{T_i} \zeta_\ell^2$ is sub-exponentially distributed with parameters $(4T_i, 4)$ and has expected value $T_i$. 
    Applying the two-sided version of~\citet[Proposition 2.9]{wainwright2019high} with $t= cT_i$, $c\in[0, 1]$, we obtain
    \begin{align}\label{eq:proofInterBurnIn}
        \Prob\left[\vert v_k^\top Y^\top Y v_k- T_i \vert \ge cT_i  \right] 
        &= \Prob\left[\Big\vert \sum_{\ell=1}^{T_i} \zeta_\ell^2 - T_i \Big\vert \geq cT_i \right] 
        \nonumber\\ 
        &\le  \frac{\delta}{9^{n_x}} \coloneqq  2 e^{-\frac{c^2 T_i}{8}}.
    \end{align}
    Observing that 
    \begin{equation*}
        \vert v_k^\top Y^\top Y v_k - T_i \vert  
        \geq cT_i 
        \Leftrightarrow 
        \Big\vert \tfrac{1}{T_i} v_k^\top  Y^\top Y v_k - 1 \Big\vert 
        \geq c,
    \end{equation*} 
    union bounding over all $v_k$, and solving the right-hand side of~\eqref{eq:proofInterBurnIn} for a burn-in-time condition, we conclude
    \begin{equation}\label{eq:proofMinEvFinal}
        \left\Vert \frac{1}{T_i} Y^\top Y - I \right\Vert \le 2 c
    \end{equation}
    with probability at least $1-\delta$ if $T_i \ge \frac{8}{c^2} \log(2\cdot 9^{n_x}/\delta)$.
    Combining~\eqref{eq:proofMinEvFinal} with~\eqref{eq:implication} results in the desired result, where we require $c\le \frac12$ for a non-trivial lower bound. 
\end{pf}
\vspace*{-1.5\baselineskip}
The following general result is a direct consequence of Hoeffding's inequality and a covering argument.
\vspace*{-\baselineskip}
\begin{lem} \label{lemma:concetrSubGSum}
    Let $x^{(\ell)} \stackrel{\text{i.i.d.}}{\sim} \subGvec{\sigma_x^2}{n_x}$ and fix a failure probability $\delta \in (0,1)$. Then it holds that 
    \begin{equation*}
        \Prob\left[\max_{v\in \sph^{n_x-1}} \vert v^\top \sum_{\ell=1}^{T_i} x^{(\ell)} \vert \le \tfrac{4}{3} \sigma_x \sqrt{2 T_i \log(9^{n_x}/\delta)}\right] \ge 1- \delta. 
    \end{equation*}
\end{lem}
\vspace*{-\baselineskip}
\vspace*{-1.25\baselineskip}
\begin{pf}
    First, we define $\sigma_x y^{(\ell)} = x^{(\ell)}$ such that $y^{(\ell)} \stackrel{\text{i.i.d.}}{\sim} \subGvec{1}{n_x}$. 
    Note that 
    \begin{equation*}
        \max_{v \in \sph^{n_x-1}}\Big\vert  v^\top \sum_{\ell=1}^{T_i} x^{(\ell)} \Big\vert 
        = \sigma_x \max_{v \in \sph^{n_x-1}}\Big\vert  v^\top \sum_{\ell=1}^{T_i} y^{(\ell)} \Big\vert,
    \end{equation*}
    i.e., it suffices to analyze 
    \begin{equation}\label{eq:normSubdiag}
        \max_{v \in \sph^{n_x-1}}\Big\vert  v^\top \sum_{\ell=1}^{T_i} y^{(\ell)} \Big\vert  =  \Big\Vert \sum_{\ell=1}^{T_i} y^{(\ell)} \Big\Vert_2. 
    \end{equation}
    Since we cannot evaluate the maximum in~\eqref{eq:normSubdiag} directly, we approximate the maximum over the sphere by the maximum over an $\epsilon$-cover. 
    To this end, we define $\mathcal{N}_\epsilon$ to be an $\epsilon$-net of $\sph^{n_x-1}$. 
    Defining $v^* $ as the maximizer of \eqref{eq:normSubdiag} it follows that for some $ v_k \in \mathcal{N}_\epsilon$ that approximates $v^* = \argmax_{v \in \sph^{n_x-1}} \vert v ^\top \sum_{\ell=1}^{T_i} y^{(\ell)} \vert$, such that $\Vert v_k - v^*\Vert \le \epsilon$, where $\epsilon\in [0, 1)$, we have 
    \begin{align}
        \Big\vert v_k^\top \sum_{\ell=1}^{T_i} y^{(\ell)} \Big\vert 
        &\geq \Big\vert  {v^*}^\top \sum_{\ell=1}^{T_i} y^{(\ell)}\Big\vert 
        - \Big\vert ( v^* - v_k)^\top \sum_{\ell=1}^{T_i} y^{(\ell)} \Big\vert 
        \nonumber\\ 
        &\ge (1-\epsilon) \Big\Vert \sum_{\ell=1}^{T_i} y^{(\ell)} \Big\Vert_2. 
        \label{eq:interm1}
    \end{align}
    Choosing $\epsilon = \frac14$ yields with 
    \citet[Lemma 5.2]{vershynin2010introduction} that $\vert \mathcal{N}_\epsilon\vert \le 9^{n_x}$, and by~\eqref{eq:interm1} we obtain
    \vspace{-25pt}
    \begin{small}
    \begin{equation}\label{eq:eps-net1}
        \Big\Vert \sum_{\ell=1}^{T_i} y^{(\ell)} \Big\Vert_2 \hspace{-3pt}
        = \hspace{-5pt}\max_{ v \in \sph^{n_x-1}}\Big\vert v^\top \sum_{\ell=1}^{T_i}  y^{(\ell)} \Big\vert 
        \leq \frac{4}{3} \max_{\ell \in \mathbb{N}_{[1, 9^{n_x}]}} \Big\vert v_k^\top \sum_{\ell=1}^{T_i} y^{(\ell)} \Big\vert. 
    \end{equation}
    \end{small}
    Note that, by construction, $v_k ^\top y^{(\ell)} \sim \subG{1}$ for any integer $\ell \in \mathbb{N}_{[1, 9^{n_x}]}$ since $y^{(\ell)} \sim \subGvec{1}{n_x}$. 
    Thus, by applying Hoeffding's inequality for each $v_k$ we obtain
    \begin{equation*}
        \Prob\left[\Big\vert  \sum_{\ell=1}^{T_i} v_k^\top y^{(\ell)} \Big\vert 
        \geq \sqrt{2 T_i \log(9^{n_x}/\delta)}\right] \le \frac{\delta}{9^{n_x}}.
    \end{equation*}
    Taking the union bounding over all $v_k \in \mathcal{N}_\epsilon$ yields 
    \begin{equation*}
        \Prob\left[\max_{\ell \in \mathbb{N}_{[1, 9^{n_x}]}} \Big\vert \sum_{\ell=1}^{T_i} v_k^\top y^{(\ell)} \Big\vert \leq \sqrt{2 T_i \log(9^{n_x}/\delta)}\right] \ge 1- \delta.
    \end{equation*}
    By~\eqref{eq:eps-net1} and~\eqref{eq:normSubdiag}, we deduce 
    \begin{equation*}
        \Prob\left[\max_{v\in \sph^{n_x-1}} \Big\vert v^\top \sum_{\ell=1}^{T_i} x^{(\ell)} \Big\vert \leq \frac{4}{3} \sigma_x\sqrt{2 T_i \log(9^{n_x}/\delta)}\right] \ge 1- \delta
    \end{equation*}
    which concludes the proof.
\end{pf}
\vspace*{-0.5\baselineskip}
%
%
\vspace*{-\baselineskip}
\section{Proof of Lemma~\ref{lemma:distOflifting}}\label{app:proofdistOfLifting}
\vspace*{-0.75\baselineskip}
First, we show that if $x \simiid \uniform{a}$ then $\sin(x)$ is a zero-mean random variable. 
In particular, 
\begin{equation*}
    \Expect[\sin(x)]
    = \int_{-\infty}^{\infty} \sin(x) f_X(x) \mathrm{d}x
    =  \frac{1}{2a} \int_{-a}^a \sin(x)\mathrm{d}x 
    = 0.
\end{equation*} 
Further, observe that $\sin(x) \in [-1, 1]$ for all $x\in\R$ and let $v \in \sph^1$ and $\lambda \in \R$.
Now, let $\varepsilon$ be an independent Rademacher variable.\footnote{A Rademacher variable $\varepsilon$ takes the values
$\{-1, 1\}$ with equal probability.}
Note that since $v^\top \xi $ is symmetric, $v^\top \xi $ and  $\varepsilon v^\top \xi $ share the same distribution. 
Hence, 
\begin{multline*}
    \Expect\left[ e^{\lambda v^\top \xi}\right] 
    = \Expect\left[\Expect_\varepsilon\left[ e^{\varepsilon \lambda v^\top \xi}\right]\right]  \\
    \le \Expect \left[e^{\frac{\lambda^2 (v^\top \xi)^2}{2}}\right] 
    = \Expect \left[e^{\frac{\lambda^2 (v_1 x + v_2 \sin(x))^2}{2}}\right],
\end{multline*} 
where the inequality follows from~\citet[Example 2.3]{wainwright2019high}. 
Since the variance proxy needs to hold for all $v \in \sph^1$, we consider 
\begin{equation*}
    \max_{v\in \sph^1} v_1^2 x^2 + v_2^2 \sin^2(x) + 2v_1 v_2 x \sin(x). 
\end{equation*}
In the following, we crudely bound this by 
\begin{equation*}
    \max_{v\in \sph^1} v_1^2 a^2 + v_2^2 + 2 a = \max_{v_1\in [-1, 1]} v_1^2(a^2 -1) + 2a+ 1.
\end{equation*}
Clearly, this yields 
\begin{equation*}
    \Expect\left[e^{\lambda v^\top \xi}\right] \le \begin{cases}
        e^{\frac{\lambda^2}{2}(2a+1)}, &\text{if } a \in (0, 1], \\
        e^{\frac{\lambda^2}{2}(a^2 + 2a)}, &\text{if } a \in (1, \infty),
    \end{cases}
\end{equation*}
which concludes the proof.
\null\hfill$\square$
%
%
\section{Proof of Proposition~\ref{prop:QDelta-sample-complexity}}\label{sec:appendix:proofControlBoundsIndividual}
\vspace*{-0.75\baselineskip}
We prove Proposition~\ref{prop:QDelta-sample-complexity} by deriving the quadratic bound in~\eqref{eq:proportional-bound-residual} from the individual identification error bounds in~\eqref{eq:errorBoundIndiv}.
To this end, recall the residual $r(x,u)$ in~\eqref{eq:residual-sampling-error} and observe that we can rewrite it as%
\vspace*{-2.25\baselineskip}
\small
\begin{equation}\label{eq:residual-rewritten}
    r(x,u) 
    = (1 - \sum_{i=1}^{n_u}[u]_i) \Delta A x
    + \Delta B_0 u_i
    + \sum_{i=1}^{n_u}[u]_i \Delta B_i x,
\vspace*{-0.75\baselineskip}
\end{equation}
\normalsize
where $\Delta A = A-\hat{A}$, $\Delta B_0 = B_0-\hat{B}_0$, $\Delta B_i = B_i-\hat{B}_i$.
Then, the identification error bounds~\eqref{eq:errorBoundIndiv} yield%
\vspace*{-2.25\baselineskip}
\small%
\begin{align*}
    \|r(x,u)\|_2
    &\leq |1-\sum_{i=1}^{n_u}[u]_i| \|\Delta A\|_2 \|x\| 
    + \sqrt{\sum_{i=1}^{n_u}\|[\Delta B_0]_i\|_2^2} \|u\|
    \nonumber\\ &\qquad
    + \sum_{i=1}^{n_u}|[u]_i| \|\Delta B_i\|_2 \|x\| 
    \\
    &\leq \left[\max_{u\in\mathbb{U}} |1-\sum_{i=1}^{n_u}[u]_i|\right] \varepsilon_A \|x\| 
    + \sqrt{\sum_{i=1}^{n_u}\varepsilon_{[B_0]_i}^2} \|u\|
    \nonumber\\ &\qquad
    + \left[\max_{u\in\mathbb{U}}\sum_{i=1}^{n_u}|[u]_i| \varepsilon_{B_i}\right] \|x\|
\\[-1.5\baselineskip]
\end{align*}
\normalsize%
with probability at least $1-\delta$.
Recall $c_x$, $c_u$ in~\eqref{eq:constants-cx-cu} and observe with probability at least $1-\delta$ that
\vspace*{-0.35\baselineskip}
\begin{equation*}
    \|r(x,u)\|_2^2 
    \leq (c_x\|x\|_2 + c_u\|u\|_2)^2 \leq 2 c_x^2 \|x\|_2^2 + 2 c_u^2 \|u\|_2^2 
    .
\vspace*{-0.5\baselineskip}
\end{equation*} 
This ensures the quadratic bound in~\eqref{eq:proportional-bound-residual} with probability at least $1-\delta$ for $Q_\Delta$ as in~\eqref{eq:QDelta-sample-complexity}.
\null\hfill$\square$
%
%
\vspace*{-0.85\baselineskip}
\section{Proof of Proposition~\ref{prop:QDelta-data-ellipsoidal}}\label{sec:appendix:proofControlBoundsEllipsoidal}
\vspace*{-0.75\baselineskip}
In the following, we prove Proposition~\ref{prop:QDelta-data-ellipsoidal} by deriving the quadratic bound in~\eqref{eq:proportional-bound-residual} from the ellipsoidal identification error bounds in~\eqref{eq:ellipBoundsTotal}.
Based on the representation of the remainder in~\eqref{eq:residual-rewritten}, we deduce with probability at least $1-\delta$%
\vspace*{-2.25\baselineskip}
\small%
\begin{align*}
    &\|r(x,u)\|_2^2
    \\
    &=
    \left(\star\right)^\top
    \Big(
        (1 - \sum_{i=1}^{n_u}[u]_i)\Delta A x
        + \Delta B_0 u
        + \sum_{i=1}^{n_u}[u]_i \Delta B_i x
    \Big) 
    \\
    &= 
    \left(\star\right)^\top
    \Big(
        (1 - \sum_{i=1}^{n_u}[u]_i)\Delta A x
        + \sum_{i=1}^{n_u}\left[\begin{smallmatrix} 
            \Delta B_i^\top \\
            [\Delta B_0]_i^\top
        \end{smallmatrix}\right]^\top 
        \left[\begin{smallmatrix}
            [u]_i x \\ [u]_i
        \end{smallmatrix}\right]
    \Big) 
    \\
    &\leq 
    x^\top (n_u+1) (1 - \sum_{i=1}^{n_u}[u]_i)^2 {\Delta A^\top \Delta A} x
    \\ &\quad  
    + \sum_{i=1}^{n_u} (n_u+1) 
    \left[\star\right]^\top
    \left[\star\right]
    \left[\begin{smallmatrix} 
        \Delta B_i^\top \\
        [\Delta B_0]_i^\top
    \end{smallmatrix}\right]^\top 
    \left[\begin{smallmatrix}
        [u]_i x \\ [u]_i
    \end{smallmatrix}\right]
    \\
    &\leq 
    (n_u+1) \Big(
        (1 - \sum_{i=1}^{n_u}[u]_i)^2 x^\top \mathcal{E}_A x
        + \sum_{i=1}^{n_u} 
        \left[\star\right]^\top
        \mathcal{E}_{B_i}
        \left[\begin{smallmatrix}
            [u]_i x \\ [u]_i
        \end{smallmatrix}\right]
    \Big),
\\[-1.5\baselineskip]
\end{align*}%
\normalsize%
where we exploit the binomial expansion for the penultimate inequality and~\eqref{eq:ellipBoundsTotal} for the last inequality.
Hence,%
\small%
\vspace*{-2.5\baselineskip}
\begin{align*}
    \|r(x,u)\|_2^2
    &\leq  
    x^\top (n_u+1) \max_{u\in\mathbb{U}}|1 - \sum_{i=1}^{n_u}[u]_i|^2 \mathcal{E}_A x
    \\ 
    &\quad+ (n_u+1)
    \left[\star\right]^\top
    \left[\star\right]^\top
    \tilde{\mathcal{E}}_B
    \left[\begin{smallmatrix}
        (u \otimes I_{n_x}) & 0 \\ 0 & I_{n_u}
    \end{smallmatrix}\right]
    \left[\begin{smallmatrix}
        x \\ u
    \end{smallmatrix}\right].
\\[-1.5\baselineskip]
\end{align*}%
\normalsize%
with probability at least $1-\delta$.
Then, defining $\hat{\mathcal{E}}_B$ as in~\eqref{eq:tildeE_B-ellipsoidal} yields the $Q_\Delta$ in~\eqref{eq:QDelta-data-ellipsoidal}.
\null\hfill$\square$

\vspace*{-\baselineskip}
\bgroup
\setlength{\columnsep}{4pt}
\small
\begin{wrapfigure}[8]{l}{0.8in}
	\includegraphics[width=0.8in,height=1.05in,clip,keepaspectratio]{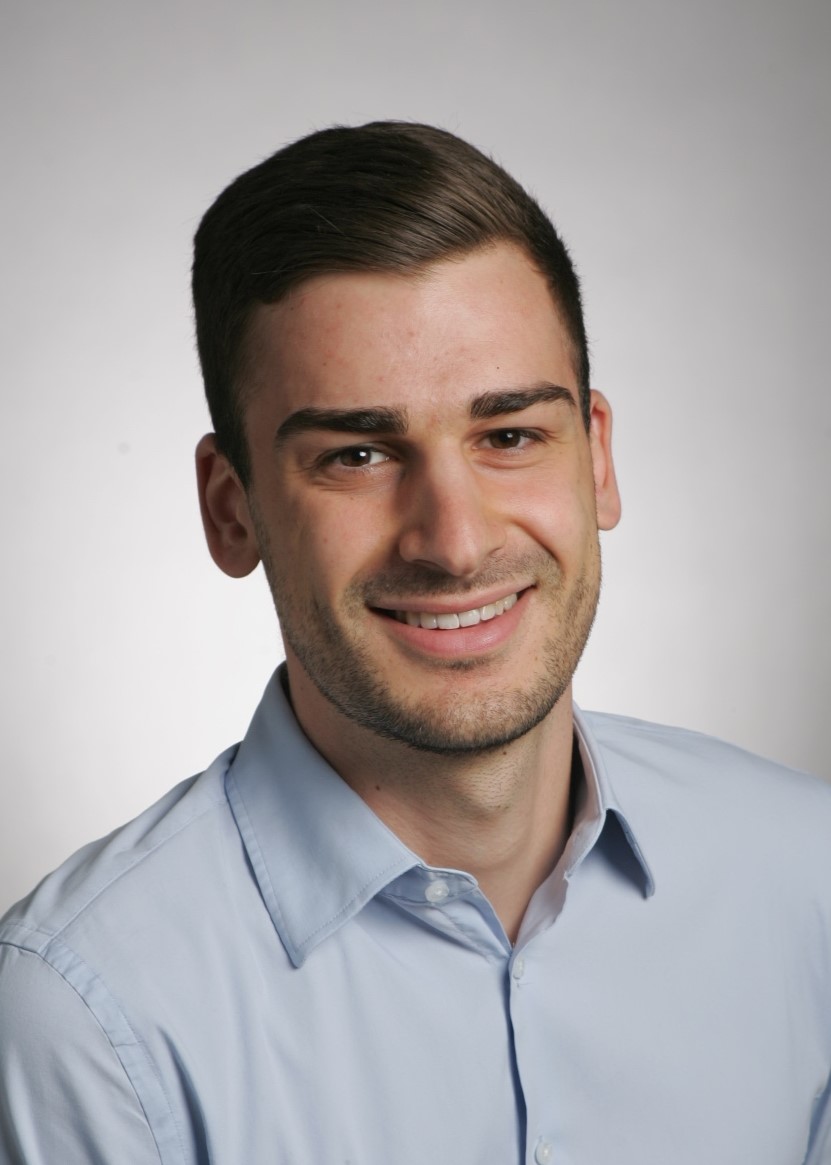}
\end{wrapfigure}
\noindent
\textbf{Nicolas Chatzikiriakos} received his M.Sc. and B.Sc. degrees in Engineering Cybernetics from the University of Stuttgart. Since 2023, he has been a Research and Teaching Assistant with the Institute for Systems Theory and Automatic Control and a member of the Graduate School Simulation Technology, both at the University of Stuttgart. His research concentrates on data-driven control through statistical learning theory, focusing on uncertainty quantification and sample efficiency in system identification. In 2025, he served as a visiting researcher at the Allen School of Computer Science \& Engineering, University of Washington, Seattle, WA, USA.

\begin{wrapfigure}[8]{l}{0.8in}
	\vspace{-8pt}
	\includegraphics[width=0.8in,height=1.05in,clip,keepaspectratio]{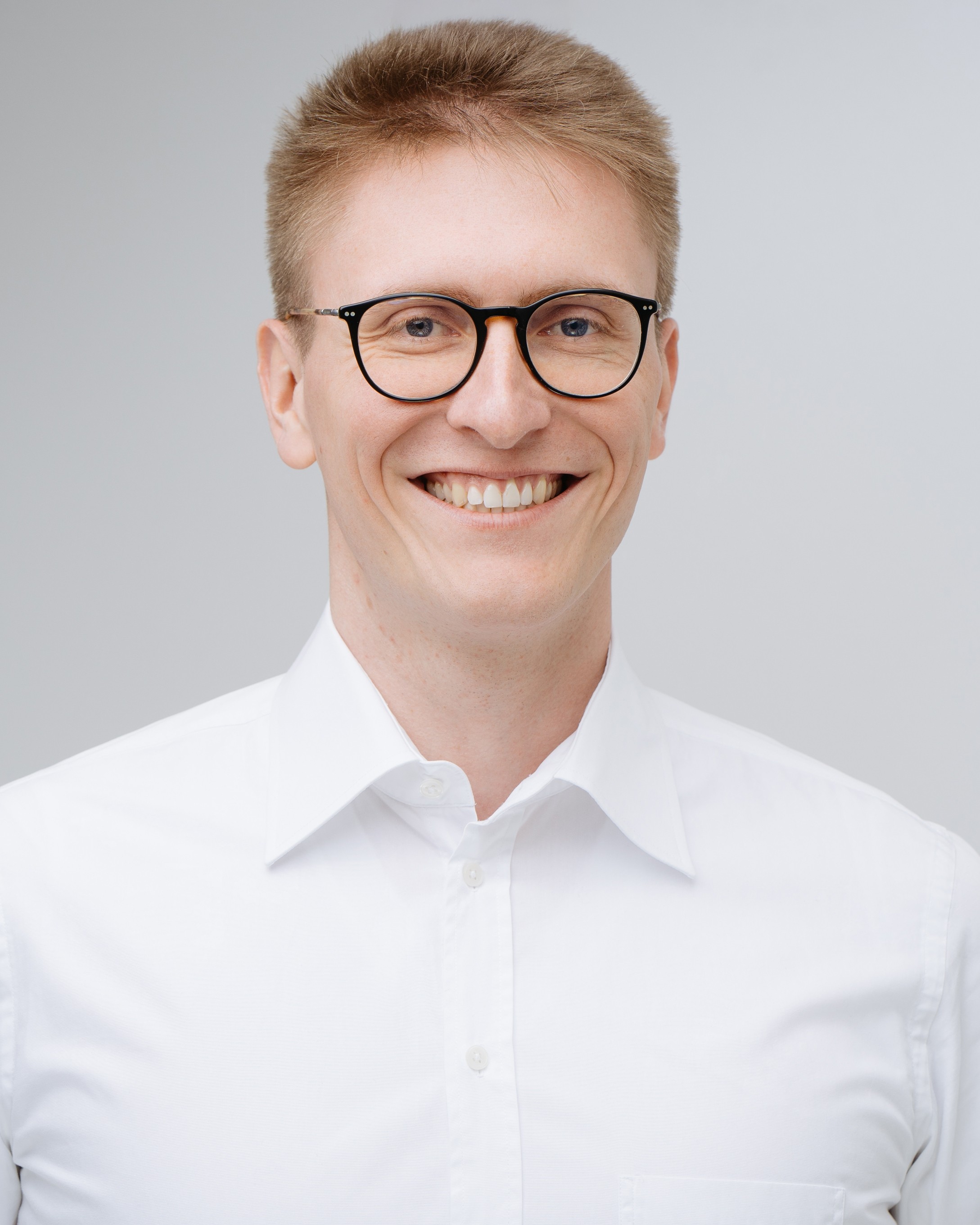}
	\vspace{-12pt}
\end{wrapfigure}
\noindent
\textbf{Robin Str\"asser} received a master’s degree in Simulation Technology from the University of Stuttgart, Germany, in 2020. Since 2020, he has been a Research and Teaching Assistant with the Institute for Systems Theory and Automatic Control and a member of the Graduate School Simulation Technology at the University of Stuttgart. His research interests include data-driven system analysis and control, with a focus on nonlinear systems. Robin Strässer received the Best Poster Award at the International Conference on Data-Integrated Simulation Science (SimTech2023) and the Best Paper Award at the European Robotics Forum (ERF2025).

\begin{wrapfigure}[8]{l}{0.8in}
	\vspace{-8pt}
	\includegraphics[width=0.8in,height=1.05in,clip,keepaspectratio]{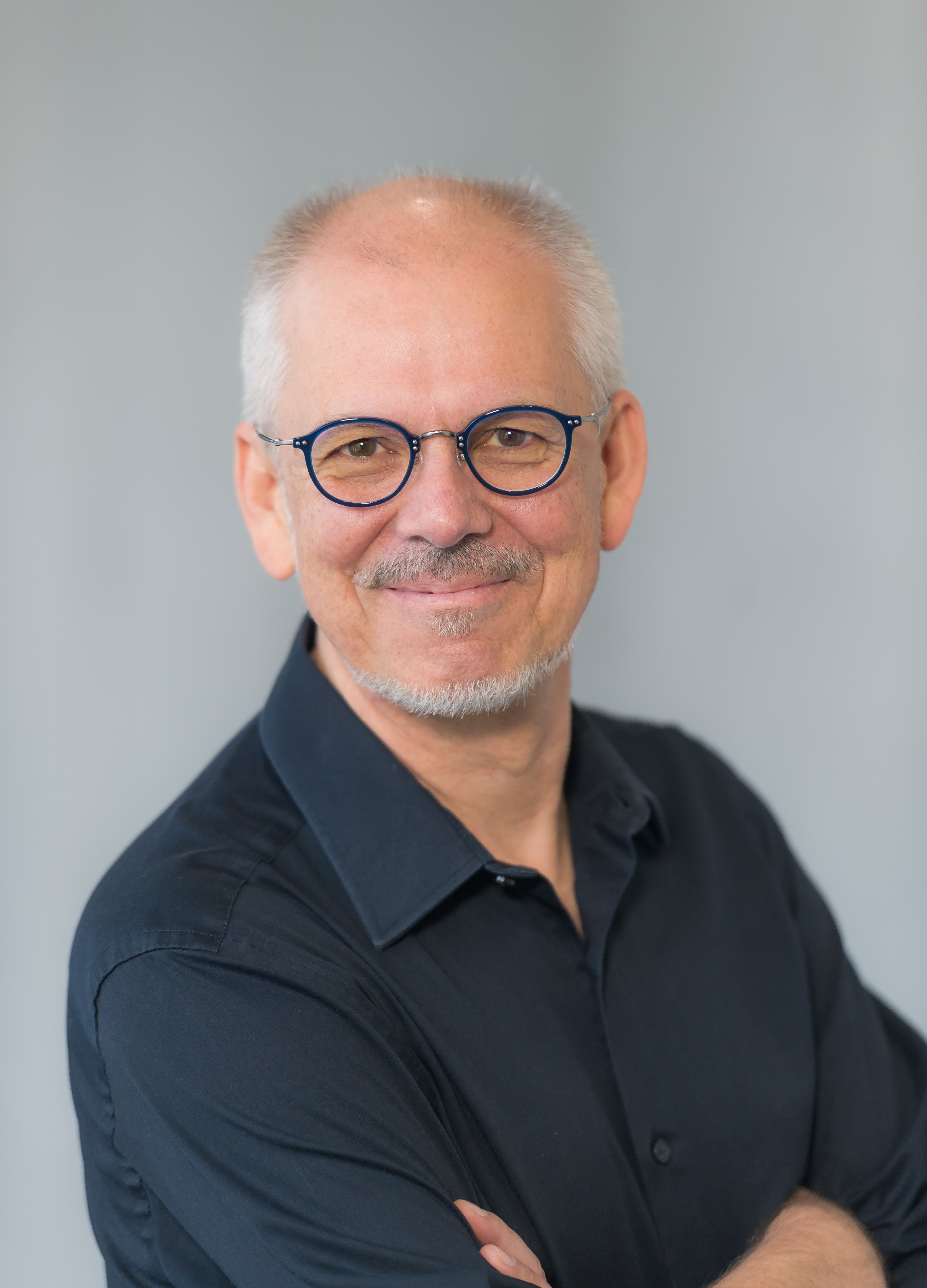}
	\vspace{-12pt}
\end{wrapfigure}
\noindent
\textbf{Frank Allg\"ower} studied engineering cybernetics and applied mathematics in Stuttgart and with the University of California, Los Angeles (UCLA), CA, USA, respectively, and received the Ph.D. degree from the University of Stuttgart, Stuttgart, Germany. Since 1999, he has been the Director of the Institute for Systems Theory and Automatic Control and a professor with the University of Stuttgart. His research interests include predictive control, data-based control, networked control, cooperative control, and nonlinear control with application to a wide range of fields including systems biology. Dr. Allgöwer was the President of the International Federation of Automatic Control (IFAC) in 2017–2020 and the Vice President of the German Research Foundation DFG in 2012–2020.

\begin{wrapfigure}[8]{l}{0.8in}
	\vspace{-8pt}
	\includegraphics[width=0.8in,height=1.05in,clip,keepaspectratio]{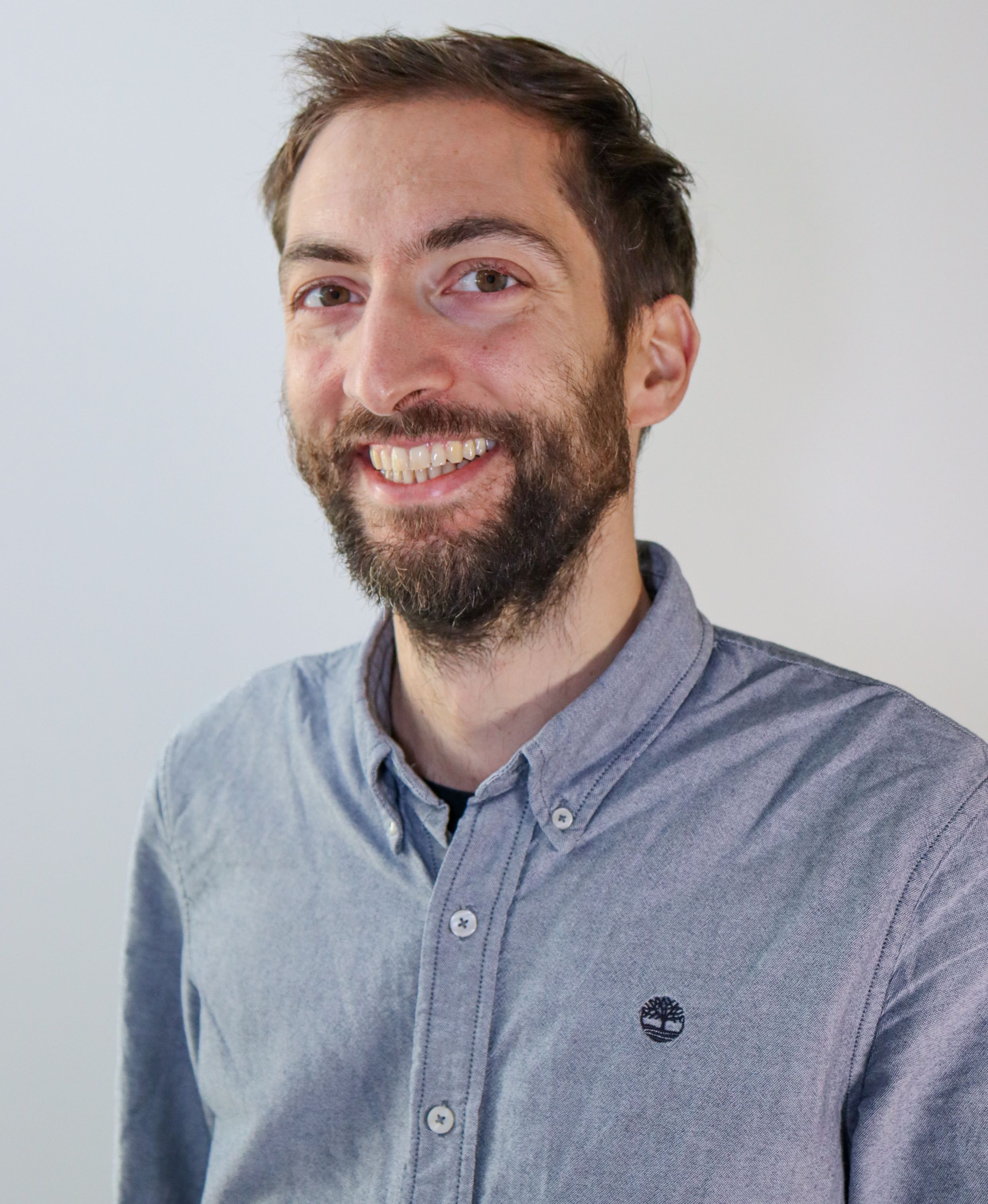}
	\vspace{-12pt}
\end{wrapfigure}
\noindent
\textbf{Andrea Iannelli} is an Assistant Professor in the Institute for Systems Theory and Automatic Control at the University of Stuttgart. 
He completed his B.Sc. and M.Sc. degrees in Aerospace Engineering at the University of Pisa and received his PhD from the University of Bristol. 
He was also a postdoctoral researcher in the Automatic Control Laboratory at ETH Zurich. 
His main research interests are centered around robust and adaptive control, uncertainty quantification, and sequential decision-making. 
He serves the community as Associated Editor for the International Journal of Robust and Nonlinear Control and as IPC member of international conferences in the areas of control, optimization, and learning.

\egroup

\end{document}